\documentclass[showpacs,showkeys,preprintnumbers]{revtex4}%
\usepackage{amsfonts}
\usepackage{amsmath}
\usepackage{graphicx}
\usepackage{dcolumn}
\usepackage{bm}
\usepackage{amssymb}%
\begin{document}
\title{Inhomogeneous condensation in effective models for QCD using the finite-mode approach}
\author{Achim Heinz$^{a}$, Francesco Giacosa$^{a,b}$, Marc Wagner$^{a}$, Dirk
H.\ Rischke$^{a}$}
\affiliation{$^{a}$~Goethe-Universit\"at Frankfurt am Main, Institut f\"ur Theoretische
Physik, Max-von-Laue-Stra{\ss }e 1, D-60438 Frankfurt am Main, Germany}
\affiliation{$^{b}$~Institute of Physics, Jan Kochanowski University, ul.\ Swietokrzyska
15, 25-406 Kielce, Poland}

\begin{abstract}
We use a numerical method, the finite-mode approach, to study inhomogeneous
condensation in effective models for QCD in a general framework. Former
limitations of considering a specific ansatz for the spatial dependence of the
condensate are overcome. Different error sources are analyzed and strategies
to minimize or eliminate them are outlined. The analytically known results for
$1+1$ dimensional models (such as the Gross-Neveu model and extensions of it)
are correctly reproduced using the finite-mode approach. Moreover, the NJL
model in $3+1$ dimensions is investigated and its phase diagram is determined
with particular focus on the inhomogeneous phase at high density.

\end{abstract}

\maketitle


\section{Introduction}

Quantum Chromodynamics (QCD) cannot be solved analytically at low energies.
However, several aspects of QCD can be understood by using effective models
which exhibit the same symmetries as QCD, most notably chiral symmetry. Some
models utilize exclusively hadronic degrees of freedom [such as chiral
$\sigma$ models
\cite{Meissner:1987ge,Ko:1994en,Urban:2001ru,Gallas:2009qp,Parganlija:2012fy}%
], while others feature constituent quarks [such as the Nambu--Jona-Lasinio
(NJL) model
\cite{Nambu:1961tp,Nambu:1961fr,Klevansky:1992qe,Vogl:1991qt,Hatsuda:1994pi,
Buballa:2003qv} and the Gross-Neveu (GN) model \cite{Gross:1974jv}]. A model
with both hadronic and quark degrees of freedom has also been discussed
\cite{vanBeveren:2002mc,Schaefer:2007pw}.

All these effective descriptions of QCD include the spontaneous breaking of
chiral symmetry, which implies the emergence of a chiral condensate at low
temperatures and densities denoted as $\sigma$. This quantity is represented
by a nonzero expectation value of a scalar-isoscalar mesonic field in hadronic
chiral models or, equivalently, by the quark-antiquark expectation value
$\langle\bar{\psi} \psi\rangle$ in quark-based models.

The chiral condensate is, in general, a function of space, $\sigma=\sigma
(\vec{x})$. In principle, the determination of $\sigma(\vec{x})$ is
straightforward: one has to find the field configuration which minimizes the
effective action at a given temperature and density. In practice, this task
is, however, very difficult. This is why $\sigma$ is often assumed to be
spatially constant. This assumption is usually valid in the vacuum and at low
densities, but not anymore at high densities. One of the simplest non-constant
field configurations is the so-called chiral density wave (CDW) which, in
chiral hadronic models, corresponds to a one-dimensional condensate of the
form $\sigma(x_{3})=\phi\cos(px_{3})$ together with pion condensation,
$\pi^{0}(x_{3})=\phi\sin(px_{3})$. Various studies have found that the CDW is
favorable compared to a constant condensate at sufficiently high densities
\cite{Baym:1973zk,Sawyer:1973fv,Campbell:1974qt,Campbell:1974qu,Dashen:1974ff,
Baym:1975tm,Migdal:1978az,Baym:1980jx,Dickhoff:1981wk,Kolehmainen:1982jn,
Baym:1982ca,Akmal:1997ft}. Interestingly, a CDW has recently also been
obtained within the extended Linear Sigma Model \cite{Heinz:2013hza}, which is
a general chiral hadronic model with (axial-)vector degrees of freedom
\cite{Gallas:2009qp,Parganlija:2012fy}. Moreover, inhomogeneous phases were
also investigated in
Refs.\ \cite{Broniowski:1989tw,Sadzikowski:2000ap,Casalbuoni:2005zp,
Nickel:2009wj,Partyka:2010gk,Broniowski:2011ef,Carignano:2012sx,Ma:2013vga,
Buballa:2014tba,Kojo:2009ha,Kojo:2011cn,Harada:2015lma,Carignano:2010ac,Muller:2013tya}
in the framework of the NJL model as well as in the quark-meson model and the
skyrmion model.

A general method to determine space-dependent condensates at non-zero
temperature and density has not yet been established. There are a few
exploratory studies of such methods in the context of the $1+1$ dimensional GN
model, using either a lattice regularization \cite{deForcrand:2006zz} of the
effective action or an expansion in terms of plane waves or hat-like localized
basis functions \cite{Wagner:2007he,Wagner:2007av}, but quite often one uses a
specific Ansatz \cite{Nickel:2009wj,Abuki:2011pf}. More evolved models
including three spatial dimensions and two-dimensional variations of the
condensates and corresponding general methods, which are also based on
expansions of fermionic fields and condensates in terms of plane waves, have
been discussed theoretically in
Refs.\ \cite{Nickel:2008ng,Carignano:2012sx,Cao:2015rea}. In practice,
however, due to limited computational resources, investigations have again
been limited to specific Ans\"{a}tze, where only a selected set of Fourier
modes is considered, e.g.\ variations in only a single spatial dimension. In
this respect models for which analytic inhomogeneous solutions are known are
extremely interesting. This is the case for the $1+1$ dimensional GN model
\cite{Thies:2003kk,Basar:2009fg,Ebert:2014woa,Kojo:2014fxa,Braun:2014fga},
where a soliton-like solution for the spatial dependence of the condensate is
found, which is mathematically represented by a Jacobi elliptic function
\cite{Thies:2003kk,Basar:2009fg}. Further interesting $1+1$ dimensional models
for which inhomogeneous phases have analytically been determined are
extensions of the GN model: the chiral Gross-Neveu ($\chi$GN) model
\cite{Salcedo:1990rw,Dashen:1975xh}, which has a continuous chiral symmetry,
and the two-flavor NJL$_{2}$ model \cite{Ebert:2009mr}. These $1+1$
dimensional models are relevant, because at high densities QCD effectively
reduces from $3+1$ to $1+1$ dimensions
\cite{McLerran:2007qj,McLerran:2008ua,Kojo:2009ha}.

Thus, while the existence of inhomogeneous phases has been verified by several
different approaches, it is highly desirable to develop a general and reliable
numerical method to study inhomogeneous condensation, which does not require a
specific ansatz for the spatial dependence of the condensate. This is the aim
of the present work. We adapt and extend techniques introduced and explored in
Refs.\ \cite{Wagner:2007he,Wagner:2007av}. We first test the validity and
reliability of the resulting method, the finite-mode approach, by applying it
to $1+1$ dimensional models, the GN, the $\chi$GN, and the NJL$_{2}$ model. We
correctly reproduce both soliton-like and CDW modulations without supplying
any specific ansatz.

Then we apply the finite-mode approach to study the phase structure of the
$3+1$ dimensional NJL model. Recent findings \cite{Nickel:2009wj} concerning
one-dimensional modulations are confirmed. In addition, we determine the shape
of the so-called inhomogeneous ``continent'' at high density
\cite{Carignano:2011gr,Carignano:2014jla}: in agreement with these works, the
phase boundary between chirally restored and inhomogeneous phase first
increases with temperature. However, for larger chemical potential $\mu$ it
decreases. Thus, the inhomogeneous phase exhibits a shape which is
surprisingly similar to that of the crystal phase of the GN model.

The paper is organized as follows. In Sec.~\ref{SEC2} quark-based effective
models for QCD in $1+1$ and $3+1$ dimensions are introduced. In
Sec.~\ref{SEC_GN}, Sec.~\ref{SEC_chiGN}, Sec.~\ref{SEC_NJL2}, and
Sec.~\ref{SEC_NJL} the phase diagrams of these models are investigated
numerically using the finite-mode approach, with particular focus on
inhomogeneous condensation. Finally, we present conclusions and an outlook in
Sec.~\ref{SEC4}.


\section{\label{SEC2}Quark-based effective models}

In this section we introduce the Lagrangians of the models that we use to
investigate inhomogeneous condensation. We start with $1+1$ dimensional models
and then turn to the $3+1$ dimensional NJL model.


\subsection{\label{SEC_1D}$1+1$ dimensions: the GN model and its extensions}


\subsubsection*{GN model}

The GN model \cite{Gross:1974jv,Wolff:1985av,Thies:2003kk,Basar:2009fg} is a
fermionic model that contains only a single quark flavor. In the large-$N$
limit (where $N$ is the number of colors) it exhibits QCD-like features such
as asymptotic freedom, dynamical chiral symmetry breaking and its restoration,
dimensional transmutation, and meson and baryon bound states
\cite{tHooft:1973jz,Witten:1979kh,Lebed:1998st,Schnetz:2005ih}. The Lagrangian
of the GN model in Euclidean space is
\begin{equation}
\label{lgnvac}\mathcal{L}_{\text{GN}} = \sum_{j=1}^{N} \bar{\psi}_{j}
\Big(\gamma_{\mu}\partial_{\mu}+ m_{0}\Big) \psi_{j} - \frac{g^{2}}{2}
\bigg(\sum_{j=1}^{N} \bar{\psi}_{j} \psi_{j}\bigg)^{2} \; ,
\end{equation}
with $\gamma_{0} = \sigma_{1}$ and $\gamma_{1} = \sigma_{3}$ implying
$\gamma_{\mu}= \gamma_{\mu}^{\dagger}= \gamma_{\mu}^{\ast}$ and $\{
\gamma_{\mu}, \gamma_{\nu}\} = 2 \delta_{\mu\nu}$. Chiral symmetry is realized
in a discrete way, $\psi_{j} \rightarrow\gamma_{5} \psi_{j}$. The term
proportional to $m_{0}$ breaks chiral symmetry explicitly (it is analogous to
a quark mass term). Therefore, in this work it is always set to zero, $m_{0} =
0$ (similar choices are also implemented for the other models studied by us).

Spontaneous symmetry breaking is only realized in the limit $N \rightarrow
\infty$ \cite{Gross:1974jv}, since for any finite $N$ spontaneous symmetry
breaking is excluded in $1+1$ dimensions \cite{Mermin:1966fe,Coleman:1973ci}.
The chiral condensate arises upon condensation of the scalar-isoscalar field
combination $\bar{\psi}_{j} \psi_{j}$, i.e., $\sigma= \langle\bar{\psi}_{j}
\psi_{j} \rangle$ (where a sum over $j$ is implied).

In the limit $N \rightarrow\infty$ analytic solutions for thermodynamical
quantities including inhomogeneous condensation have been found
\cite{Schnetz:2005ih} (see also the discussion in Sec.~\ref{SEC_GN}).


\subsubsection*{$\chi$GN model}

A straightforward extension of the GN model is obtained by adding a
pseudoscalar term. The Lagrangian of the $\chi$GN model is
\begin{equation}
\mathcal{L}_{\chi\text{GN}} = \sum_{j=1}^{N} \bar{\psi}_{j} \gamma_{\mu
}\partial_{\mu}\psi_{j} - \frac{g^{2}}{2} \bigg[\bigg(\sum_{j=1}^{N} \bar
{\psi}_{j} \psi_{j}\bigg)^{2} + \bigg(\sum_{j=1}^{N} \bar{\psi}_{j}
\imath\gamma_{5} \psi_{j}\bigg)^{2}\bigg] \; .
\end{equation}
This model contains a scalar field combination $\bar{\psi}_{j} \psi_{j}$,
which corresponds to a $\sigma$-like particle, and a pseudoscalar field
combination $\bar{\psi}_{j} \imath\gamma_{5} \psi_{j}$, which corresponds to
an $\eta$-like particle. It is invariant under continuous $U_{A}(1)$ chiral
symmetry transformations, $\psi_{j} \rightarrow e^{\imath\theta\gamma_{5}}
\psi_{j}$, and has certain similarities to one-flavor QCD (when the chiral
anomaly is excluded).

The $\chi$GN model is particularly interesting, because both the scalar and
the pseudoscalar field configurations condense, when the temperature exceeds a
critical value. The ground state is then a CDW
\cite{Schon:2000he,Schon:2000qy}.


\subsubsection*{NJL$_{2}$ model}

A further extension of the GN model is obtained by considering, in addition to
the scalar-isoscalar field combination, three pion-like field combinations. In
this respect the model is similar to two-flavor QCD. The Lagrangian of this
so-called NJL$_{2}$ model is
\begin{equation}
\mathcal{L}_{\text{NJL}_{2}} = \sum_{f=1}^{2} \sum_{j=1}^{N} \bar{\psi}_{j,f}
\gamma_{\mu}\partial_{\mu}\psi_{j,f} - \frac{g^{2}}{2} \sum_{f=1}^{2}
\bigg[\bigg(\sum_{j=1}^{N} \bar{\psi}_{j,f} \psi_{j,f}\bigg)^{2} +
\bigg(\sum_{j=1}^{N} \bar{\psi}_{j,f} \vec{\tau} \imath\gamma_{5} \psi
_{j,f}\bigg)^{2}\bigg]\; ,
\end{equation}
where $f=1,2$ is the flavor index. The model is invariant under chiral
symmetry transformations $SU_{L}(2) \times SU_{R}(2)$,
\begin{equation}
\label{EQN560}\psi_{j,L} \rightarrow U_{L} \psi_{j,L} \quad, \quad\psi_{j,R}
\rightarrow U_{R} \psi_{j,R} \; ,
\end{equation}
with
\begin{equation}
\label{EQN560_}\psi_{j} = \left(
\begin{array}
[c]{c}%
\psi_{j,1}\\
\psi_{j,2}%
\end{array}
\right)  \quad, \quad\psi_{j,L} = P_{L} \psi_{j} \quad, \quad\psi_{j,R} =
P_{R} \psi_{j} \; ,
\end{equation}
and $P_{L}$ and $P_{R}$ are projectors onto left- and right-handed components, respectively.

In contrast to the $\chi$GN model the ground state is not a CDW. Using the
finite-mode approach we find that the phase diagram coincides with that of the
GN model (cf.\ Sec.\ \ref{SEC_NJL2}).

\subsection{\label{SEC776}$3+1$ dimensions: the NJL model}

The NJL model in $3+1$ dimensions is one of the most famous effective chiral
approaches to QCD. It has been extensively used in the vacuum and at non-zero
temperature and density to study the spontaneous breaking of chiral symmetry
and its restoration [cf.\ e.g.\ Refs.\ \cite{Bringoltz:2005rr,Panero:2009tv}].
The Lagrangian (in the chiral limit for $N$ colors and two flavors) is
\cite{Hatsuda:1994pi,Klevansky:1992qe}
\begin{equation}
\mathcal{L}_{\text{NJL}}=\sum_{f=1}^{2}\sum_{j=1}^{N}\bar{\psi}_{j,f}%
\gamma_{\mu}\partial_{\mu}\psi_{j,f}-\frac{3G}{N}\sum_{f=1}^{2}%
\bigg[\bigg(\sum_{j=1}^{N}\bar{\psi}_{j,f}\psi_{j,f}\bigg)^{2}+\bigg(\sum
_{j=1}^{N}\bar{\psi}_{j,f}\vec{\tau}\imath\gamma_{5}\psi_{j,f}\bigg)^{2}%
\bigg] \; . \label{njllag}%
\end{equation}
Chiral symmetry is realized in the same way as in the NJL$_{2}$ model
[cf.\ Eqs.\ (\ref{EQN560}) and (\ref{EQN560_})].

In the vacuum, the quark field obtains an effective mass, if the coupling
constant $G$ exceeds a critical value,
\begin{equation}
\label{eq:direct_gap}m_{0}^{\ast}=-\frac{6G}{N}\sum_{f=1}^{2} \sum_{j=1}%
^{N}\langle\bar{\psi}_{j,f}\psi_{j,f}\rangle>0 \; .
\end{equation}
This effective mass is proportional to the chiral condensate in the vacuum,
i.e., $\sigma_{0} = -(N / 6 \sqrt{2} G) m_{0}^{\ast}$
\cite{Hatsuda:1994pi,Klevansky:1992qe}, where the chiral condensate is defined
according to
\begin{equation}
\label{condnjl}\sigma= \frac{1}{\sqrt{2}} \sum_{f=1}^{2} \sum_{j=1}^{N}
\langle\bar{\psi}_{j,f} \psi_{j,f} \rangle\; .
\end{equation}
In other words, the field combination which gives rise to a non-zero
condensate is again $\bar{\psi}_{j,f} \psi_{j,f}$. When restricting this
condensate to be constant, chiral symmetry restoration at high densities
occurs via a first-order phase transition
\cite{Glendenning:2000wn,Mishustin:1993ub,Papazoglou:1997uw,
Papazoglou:1998vr,Bonanno:2007kh}. However, when allowing for an inhomogeneous
condensate, the latter occurs at slightly smaller chemical potentials than the
first-order phase transition. This is in agreement with the extended Linear
Sigma Model results of Ref.\ \cite{Heinz:2013hza}.

Contrary to the $1+1$ dimensional models of Sec.~\ref{SEC_1D}, the NJL model
is not renormalizable. The equation for $\sigma$ takes the form
\begin{equation}
\sigma_{0}=-2\sqrt{2}Nm_{0}^{\ast}\, I \quad\text{ with } \quad I=i\int
\frac{d^{4}p}{(2\pi)^{4}}\frac{2}{p^{2}-m_{0}^{\ast2}}=\int\frac{d^{3}p}%
{(2\pi)^{3}}\frac{1}{\sqrt{\vec{p}^{2}+m_{0}^{\ast2}}}\; , \label{tadpole}%
\end{equation}
where the integral $I$ corresponds to a closed quark loop, i.e., to a tadpole
diagram arising from the quartic NJL interaction of Eq.\ (\ref{njllag}), which
affects the quark propagator at the resummed one-loop level in the
Hartree-Fock approximation [see
Refs.\ \cite{Klevansky:1992qe,Hatsuda:1994pi,Vogl:1991qt} for a detailed derivation].

The integral $I$ is, however, quadratically divergent. Indeed, the NJL model
is properly defined only after a regularization scheme has been chosen and a
corresponding high-energy scale enters as a new parameter. Strictly speaking,
each choice of regularization corresponds to a different version of the NJL
model. Once the regularization has been fixed, the quantity $m_{0}^{\ast}$ in
Eq.\ (\ref{eq:direct_gap}) and, as a consequence, all the relevant
thermodynamical quantities, are finite.

Especially in studies of the NJL model at nonzero density it is common to
implement a three-dimensional cutoff [see e.g.\ Ref.\ \cite{Buballa:2003qv}]
according to which
\begin{equation}
I\rightarrow I_{\Lambda}=\int_{0}^{\Lambda}\frac{d\left\vert \vec
{p}\right\vert }{2\pi^{2}}\frac{\vec{p}^{2}}{\sqrt{\vec{p}^{2}+m_{0}^{\ast2}}%
}\; .
\end{equation}
Note that the use of a four-dimensional covariant cutoff is possible for
studies of the vacuum \cite{Klevansky:1992qe,Hatsuda:1994pi}, but it is not
easy to implement at nonzero temperatures and densities.

However, a three-dimensional cutoff strongly suppresses the appearance of
inhomogeneous phases. Namely, in an inhomogeneous phase such as a CDW the
quark propagator is not diagonal and the ingoing and outgoing momenta can
differ by a full wavelength. This hardly takes place when the momentum
$\left\vert \vec{p}\right\vert $ is limited by the cutoff $\Lambda$
\cite{Buballa:pri}. Hence, in order to realize a CDW, other regularization
approaches must be used, such as the Pauli-Villars scheme
\cite{Carignano:2014jla} or the proper-time regularization scheme
\cite{Vogl:1991qt,Hatsuda:1994pi,Nakano:2004cd}.

In this work we use the Pauli-Villars approach, which is a Lorentz (and gauge)
invariant regularization procedure \cite{Klevansky:1992qe,Itzykson:2006wn}. It
amounts to introduce $N_{\text{PV}}$ additional fictitious heavy fermions with
mass $M_{k}$ in such a way that the tadpole integral of Eq.\ (\ref{tadpole})
is modified according to
\begin{equation}
I\rightarrow I_{\text{PV}}=\int\frac{\vec{p}^{2}d\left\vert \vec{p}\right\vert
}{2\pi^{2}}\bigg( \frac{1}{\sqrt{\vec{p}^{2}+m_{0}^{\ast2}}}+\sum
_{k=1}^{N_{PV}}\frac{C_{k}}{\sqrt{\vec{p}^{2}+M_{k}^{2}}}\bigg) \; ,
\label{ipv}%
\end{equation}
with the masses given by
\begin{equation}
M_{k}^{2}=m_{0}^{\ast2}+\alpha_{k}\Lambda_{\text{PV}}^{2}\; ,
\end{equation}
where $\Lambda_{\text{PV}}$ is the so-called Pauli-Villars high-energy scale.
The constants $C_{k}$ and $\alpha_{k}$ are real dimensionless numbers, which
are chosen in such a way that $I_{PV}$ is finite. Let us show this explicitly
for the case $N_{\text{PV}}=2$. For large values of $\vec{p}^{2}$ the quantity
in parentheses in Eq.\ (\ref{ipv}) can be approximated by a Taylor expansion,
\begin{equation}
(\ldots)=\frac{1}{\left\vert \vec{p}\right\vert }\bigg[ 1+C_{1}+C_{2}-\frac
{1}{2}\frac{m_{0}^{\ast2}(1+C_{1}+C_{2})+\Lambda^{2}_{\mathrm{PV}} (\alpha
_{1}C_{1}+\alpha_{2}C_{2})}{\left\vert \vec{p}\right\vert ^{2}}+\mathcal{O}(
\left\vert \vec{p}\right\vert ^{-4}) \bigg]\; .
\end{equation}
Then, by requiring
\begin{equation}
1+C_{1}+C_{2}=0 \quad\text{ and } \quad\alpha_{1}C_{1}+\alpha_{2}C_{2}=0
\label{cond2}%
\end{equation}
the integrand of Eq.\ (\ref{ipv}) falls off as $\left\vert \vec{p}\right\vert
^{-3}$ and is, therefore, convergent (although it explicitly depends on the
scale $\Lambda_{\text{PV}}$). Once $I_{\text{PV}}$ is finite, the quark
condensate, the quark mass, as well as all other relevant quantities are also
finite. The conditions in Eq.\ (\ref{cond2}) are met for $\alpha_{1}=2$ and
$\alpha_{2}=1$ with $C_{1}=1$ and $C_{2}=-2$.

The procedure can be easily generalized to an arbitrary number of heavy
fermions $N_{\mathrm{PV}}$,
\begin{equation}
1+\sum_{k=1}^{N_{\mathrm{PV}}}C_{k}=0 \quad, \quad\sum_{k=1}^{N_{\mathrm{PV}}%
}\alpha_{k}C_{k}=0 \; . \label{condnpv}%
\end{equation}
For the case $N_{\mathrm{PV}}=3$ the previous equations are fulfilled by
$\alpha_{1}=1$, $\alpha_{2}=2$, $\alpha_{3}=3$ and $C_{1}=-3$, $C_{2}=3$ ,
$C_{3}=-1$. In Sec.~\ref{SEC_NJL} we will compute the phase diagram of the NJL
model with inhomogeneous condensation using the Pauli-Villars regularization
with two and three heavy fermions.


\section{\label{SEC_GN}Finite-mode regularization of the $1+1$ dimensional GN
model}

In the following we discuss the finite-mode approach in detail, in particular
its technical aspects, in the context of the $1+1$ dimensional GN model in the
large-$N$ limit [cf.\ also
Refs.\ \cite{Andrianov:1982sn,Andrianov:1983fg,Andrianov:1983qj,Wagner:2007he,
Wagner:2007av}]. We reproduce the analytically known phase diagram, which
exhibits an inhomogeneous crystal phase.


\subsection{\label{SEC559}Partition function and Euclidean action}

The partition function of the 1+1 dimensional GN model (\ref{lgnvac}) in
Euclidean spacetime is
\begin{equation}
Z=\int\bigg(\prod_{j=1}^{N}D\bar{\psi}_{j}D\psi_{j}\bigg)e^{-S_{E}[\psi
_{j},\bar{\psi}_{j}]}\;,
\end{equation}
with the action
\begin{equation}
S_{E}[\bar{\psi}_{j},\psi_{i}]=\int d^{2}x\,\bigg[\sum_{j=1}^{N}\bar{\psi}%
_{j}\Big(\gamma_{\mu}\partial_{\mu}+\gamma_{0}\mu\Big)\psi_{j}-\frac{g^{2}}%
{2}\bigg(\sum_{j=1}^{N}\bar{\psi}_{j}\psi_{j}\bigg)^{2}\bigg]\;,
\end{equation}
where $\mu$ is the chemical potential. One can get rid of the four-fermion
term by introducing a real scalar field $\sigma$,
\begin{equation}
Z=\int D\sigma\,\bigg(\prod_{j=1}^{N}D\bar{\psi}_{j}D\psi_{j}\bigg)\exp
\bigg[-\int d^{2}x\,\bigg(\frac{1}{2g^{2}}\sigma^{2}+\sum_{j=1}^{N}\bar{\psi
}_{j}Q\psi_{j}\bigg)\bigg]\;,
\end{equation}
with the Dirac operator
\begin{equation}
Q=\gamma_{\mu}\partial_{\mu}+\gamma_{0}\mu+\sigma\;.
\end{equation}
Performing the integration over the fermionic fields results in
\begin{equation}
Z=\int D\sigma\,(\det Q)^{N}\exp\bigg(-\frac{1}{2g^{2}}\sigma^{2}\bigg)\;.
\end{equation}
Since $\det Q$ is real \cite{fn}, $(\det Q)^{2}=\det(Q^{\dagger}Q)\geq0$.
Consequently, for even $N$
\begin{equation}
Z=\int D\sigma\,e^{-S_{E,\text{eff}}[\sigma]}\quad,\quad S_{E,\text{eff}%
}[\sigma]=N\int d^{2}x\,\bigg\{\frac{1}{2\lambda}\sigma^{2}-\frac{1}{2}%
\ln[\det(Q^{\dagger}Q)]\bigg\}\;,
\end{equation}
where $\lambda=Ng^{2}$. Due to numerical reasons discussed in detail in
Ref.\ \cite{Wagner:2007he}, when using the finite-mode approach it is highly
advantageous to regularize the effective action expressed in terms of
$\det(Q^{\dagger}Q)$ instead of the mathematically equivalent expression
containing $\det Q$.


\subsection{Finite-mode regularization, homogeneous condensate $\sigma=
\text{constant}$}

For numerical calculations it is convenient to work exclusively with
dimensionless quantities. Therefore, we express all dimensionful quantities in
units of $\sigma_{0}$ which is the non-vanishing value of the constant
condensate $\sigma$ at temperature $T=0$ and chemical potential $\mu=0$. The
resulting dimensionless quantities are denoted by a hat $\hat{\phantom{x}}$,
e.g.\ $\hat{x}_{\mu}= x_{\mu}\sigma_{0}$, $\hat{T} = T / \sigma_{0}$,
$\hat{\mu} = \mu/ \sigma_{0}$, $\hat{\sigma} = \sigma/ \sigma_{0}$, etc.

We consider a finite spacetime volume with temporal extension $\hat{L}%
_{0}=L_{0}\sigma_{0}$ (corresponding to the inverse temperature $1/\hat
{T}=\hat{L}_{0}$) and spatial extension $\hat{L}_{1}=L_{1}\sigma_{0}$. The
fermionic fields are expressed as superpositions of plane waves with periodic
boundary conditions in spatial direction and antiperiodic boundary conditions
in temporal direction,
\begin{align}
\hat{\psi}_{j}(\hat{x}_{0},\hat{x}_{1})  &  =\frac{\psi_{j}(x_{0},x_{1}%
)}{\sqrt{\sigma_{0}}}=\sum_{n_{0},n_{1}}\eta_{j,n_{0},n_{1}}\frac
{e^{-\imath(\hat{k}_{0}\hat{x}_{0}+\hat{k}_{1}\hat{x}_{1})}}{\sqrt{\hat{L}%
_{0}\hat{L}_{1}}}\;,\label{EQN050}\\
\hat{\bar{\psi}}_{j}(\hat{x}_{0},\hat{x}_{1})  &  =\frac{\bar{\psi}_{j}%
(x_{0},x_{1})}{\sqrt{\sigma_{0}}}=\sum_{n_{0},n_{1}}\bar{\eta}_{j,n_{0},n_{1}%
}\frac{e^{+\imath(\hat{k}_{0}\hat{x}_{0}+\hat{k}_{1}\hat{x}_{1})}}{\sqrt
{\hat{L}_{0}\hat{L}_{1}}}\;, \label{EQN051}%
\end{align}
with discrete momenta
\[
\hat{k}_{0}=\frac{2\pi}{\hat{L}_{0}}\left(  n_{0}-\frac{1}{2}\right)
\quad,\quad\hat{k}_{1}=\frac{2\pi}{\hat{L}_{1}}n_{1}\quad,\quad n_{0},n_{1}%
\in\mathbb{N}\;,
\]
where $\eta_{j,n_{0},n_{1}}$ and $\bar{\eta}_{j,n_{0},n_{1}}$ are
dimensionless Grassmann variables.

For a homogeneous condensate $\sigma=\text{constant}$, $\ln[\det(\hat
{Q}^{\dagger}\hat{Q})]$ can be expressed as a product over the modes $(\hat
{k}_{0},\hat{k}_{1})$,
\begin{align}
&  \ln[\det(\hat{Q}^{\dagger}\hat{Q})]=\ln\bigg\{\prod_{\hat{k}_{0},\hat
{k}_{1}}\det\Big[\Big(+\imath\gamma_{\mu}\hat{k}_{\mu}-\gamma_{0}\hat{\mu
}+\hat{\sigma}\Big)\Big(-\imath\gamma_{\mu}\hat{k}_{\mu}+\gamma_{0}\hat{\mu
}+\hat{\sigma}\Big)\Big]\bigg\}\nonumber\label{EQN766}\\
&  \hspace{0.5cm}=\ln\bigg\{\prod_{\hat{k}_{0},\hat{k}_{1}}\Big[(\hat{k}%
_{0}^{2}+\hat{k}_{1}^{2}+\hat{\sigma}^{2}-\hat{\mu}^{2})^{2}+(2\hat{\mu}%
\hat{k}_{0})^{2}\Big]\bigg\}\;.
\end{align}
Considering only a finite number of modes $n_{0}=-N_{0}+1,-N_{0}%
+2,\ldots,N_{0}-1,N_{0}$ and $n_{1}=-N_{1},-N_{1}+1,\ldots,N_{1}-1,N_{1}$,
i.e., introducing momentum cutoffs
\begin{equation}
\hat{k}_{0}^{\text{cut}}=\frac{2\pi}{\hat{L}_{0}}N_{0}\quad,\quad\hat{k}%
_{1}^{\text{cut}}=\frac{2\pi}{\hat{L}_{1}}\left(  N_{1}+\frac{1}{2}\right)
\;, \label{cutoffk}%
\end{equation}
(chosen to be $\pi/\hat{L}_{0,1}$ larger than the largest momenta considered)
yields the finite-mode regularized effective action,
\begin{align}
&  \frac{S_{E,\text{eff}}(\hat{\sigma})}{N}=\frac{\hat{L}_{0}\hat{L}_{1}%
\hat{\sigma}^{2}}{2\lambda}\nonumber\label{EQN002}\\
&  \hspace{1cm}-\frac{1}{2}\sum_{n_{0}=-N_{0}+1}^{N_{0}}\sum_{n_{1}=-N_{1}%
}^{N_{1}}\ln\bigg(\bigg\{\bigg[\frac{2\pi}{\hat{L}_{0}}\left(  n_{0}-\frac
{1}{2}\right)  \bigg]^{2}+\bigg(\frac{2\pi}{\hat{L}_{1}}n_{1}\bigg)^{2}%
+\hat{\sigma}^{2}-\hat{\mu}^{2}\bigg\}^{2}+\bigg[2\hat{\mu}\frac{2\pi}{\hat
{L}_{0}}\left(  n_{0}-\frac{1}{2}\right)  \bigg]^{2}\bigg)\nonumber\\
&  \hspace{0.5cm}=\frac{2\pi^{2}N_{0}(N_{1}+1/2)\hat{\sigma}^{2}}{\lambda
\hat{k}_{0}^{\text{cut}}\hat{k}_{1}^{\text{cut}}}\nonumber\\
&  \hspace{1cm}-\sum_{n_{0}=1}^{N_{0}}\sum_{n_{1}=-N_{1}}^{N_{1}}%
\ln\bigg\{\bigg[\bigg(\hat{k}_{0}^{\text{cut}}\frac{n_{0}-1/2}{N_{0}%
}\bigg)^{2}+\bigg(\hat{k}_{1}^{\text{cut}}\frac{n_{1}}{N_{1}+1/2}%
\bigg)^{2}+\hat{\sigma}^{2}-\hat{\mu}^{2}\bigg]^{2}+\bigg(2\hat{\mu}\hat
{k}_{0}^{\text{cut}}\frac{n_{0}-1/2}{N_{0}}\bigg)^{2}\bigg\}\;,
\end{align}
which is suitable for numerical evaluation. Minimizing this effective action
with respect to $\hat{\sigma}$ for various $\hat{\mu}$ and $\hat{T}$ yields
$\hat{\sigma}(\hat{\mu},\hat{T})$, i.e., the \textquotedblleft homogeneous
phase diagram of the GN model\textquotedblright\ \cite{Wolff:1985av}.

Infinite-volume continuum results are obtained in the limit $\hat{L}_{1}
\rightarrow\infty$, $\hat{k}_{0}^{\text{cut}} \rightarrow\infty$, and $\hat
{k}_{1}^{\text{cut}} \rightarrow\infty$, which implies an infinite number of
modes. Numerically one is, of course, restricted to a finite number of modes.
In the following we discuss how to determine the parameters $\lambda$, $N_{0}%
$, $N_{1}$, $\hat{k}_{0}^{\text{cut}}$, and $\hat{k}_{1}^{\text{cut}}$
[$\hat{L}_{0}$ and $\hat{L}_{1}$ are then given by Eq.\ (\ref{cutoffk})] in an
optimal way, i.e., how to obtain numerical results with a finite and rather
limited number of modes, which are nevertheless very close to infinite-volume
continuum results.


\subsection{\label{SEC033}Choosing and determining suitable parameters
$\lambda$, $N_{0}$, $N_{1}$, $\hat{k}_{0}^{\text{cut}}$, and $\hat{k}%
_{1}^{\text{cut}}$}

The condensate $\hat{\sigma}$ minimizes the effective action (\ref{EQN002}).
For $\mu=0$ it is the solution of
\begin{equation}
0=\frac{d}{d\hat{\sigma}}\frac{S_{E,\text{eff}}(\hat{\sigma})}{N}=2\hat
{\sigma}\bigg\{\frac{2\pi^{2}N_{0}(N_{1}+1/2)}{\lambda\hat{k}_{0}^{\text{cut}%
}\hat{k}_{1}^{\text{cut}}}-2\sum_{n_{0}=1}^{N_{0}}\sum_{n_{1}=-N_{1}}^{N_{1}%
}\bigg[\bigg(\hat{k}_{0}^{\text{cut}}\frac{n_{0}-1/2}{N_{0}}\bigg)^{2}%
+\bigg(\hat{k}_{1}^{\text{cut}}\frac{n_{1}}{N_{1}+1/2}\bigg)^{2}+\hat{\sigma
}^{2}\bigg]^{-1}\bigg\}\;.
\end{equation}
An obvious solution is $\hat{\sigma}=0$. It corresponds to a minimum for
$T>T_{c}$ and to a maximum for $T\leq T_{c}$. For the latter case there are
two additional solutions (corresponding to minima), which can be obtained
from
\begin{equation}
0=\frac{2\pi^{2}N_{0}(N_{1}+1/2)}{\lambda\hat{k}_{0}^{\text{cut}}\hat{k}%
_{1}^{\text{cut}}}-2\sum_{n_{0}=1}^{N_{0}}\sum_{n_{1}=-N_{1}}^{N_{1}%
}\bigg[\bigg(\hat{k}_{0}^{\text{cut}}\frac{n_{0}-1/2}{N_{0}}\bigg)^{2}%
+\bigg(\hat{k}_{1}^{\text{cut}}\frac{n_{1}}{N_{1}+1/2}\bigg)^{2}+\hat{\sigma
}^{2}\bigg]^{-1}\;. \label{EQN003}%
\end{equation}

To appropriately determine the parameters $\lambda$, $N_{0}$, $N_{1}$,
$\hat{k}_{0}^{\text{cut}}$, and $\hat{k}_{1}^{\text{cut}}$, we consider and
relate computations at $\mu= 0$ and a low temperature $T \approx0$, where
$\sigma(T) \approx\sigma_{0}$, and at $\mu= 0$ and the critical temperature $T
= T_{c}$, where $\sigma$ just vanishes, i.e., $\sigma(T-\epsilon) > 0$ and
$\sigma(T) = 0$. The parameters $\lambda$, $N_{1}$, $\hat{k}_{0}^{\text{cut}}%
$, and $\hat{k}_{1}^{\text{cut}}$ are the same for both simulations, while
$N_{0} = N_{00}$ for $T \approx0$ and $N_{0} = N_{0c} \ll N_{00}$ for $T =
T_{c}$.

The parameters $N_{0c}$, $N_{00}$, $N_{1}$, and $\hat{k}_{1}^{\text{cut}}$ can
be chosen independently. The maximum number of modes $\propto N_{00} N_{1}$
is, of course, limited by the available computer resources. Strategies for
choosing these four parameters in an optimal way, i.e., where systematic
errors due to the finite spatial extension and the finite number of modes are
minimized, are discussed in Secs.\ \ref{SEC010}, \ref{SEC650}, and
\ref{SEC011} below.

In contrast to that, $\hat{k}_{0}^{\text{cut}}$ and $\lambda$ cannot be chosen
independently: $\hat{k}_{0}^{\text{cut}}=2\pi N_{0c}\hat{T}_{c}$ [which
follows from Eq.\ (\ref{cutoffk})], i.e., $\hat{k}_{0}^{\text{cut}}$ is
related to $N_{0c}$. Since $\hat{T}_{c}$ is a priori unknown, setting $\hat
{k}_{0}^{\text{cut}}$ to an appropriate value is a non-trivial task.
Similarly, $\lambda$ depends on $N_{0c}$, $N_{1}$, $\hat{k}_{0}^{\text{cut}}$,
and $\hat{k}_{1}^{\text{cut}}$ via Eq.\ (\ref{EQN003}) at $T=T_{c}$, where
$\hat{\sigma}=0$,
\begin{equation}
\frac{\pi^{2}(N_{1}+1/2)}{\lambda\hat{k}_{0}^{\text{cut}}\hat{k}%
_{1}^{\text{cut}}}=\frac{1}{N_{0c}}\sum_{n_{0}=1}^{N_{0c}}\sum_{n_{1}=-N_{1}%
}^{N_{1}}\bigg[\bigg(\hat{k}_{0}^{\text{cut}}\frac{n_{0}-1/2}{N_{0c}%
}\bigg)^{2}+\bigg(\hat{k}_{1}^{\text{cut}}\frac{n_{1}}{N_{1}+1/2}%
\bigg)^{2}\bigg]^{-1}\;. \label{EQN004}%
\end{equation}
To determine $\hat{k}_{0}^{\text{cut}}$ (without knowing $\hat{T}_{c}$), we
consider Eq.\ (\ref{EQN003}) also for $T\approx0$, where $\hat{\sigma}%
\approx1$, i.e.,
\begin{equation}
\frac{\pi^{2}(N_{1}+1/2)}{\lambda\hat{k}_{0}^{\text{cut}}\hat{k}%
_{1}^{\text{cut}}}=\frac{1}{N_{00}}\sum_{n_{0}=1}^{N_{00}}\sum_{n_{1}=-N_{1}%
}^{N_{1}}\bigg[\bigg(\hat{k}_{0}^{\text{cut}}\frac{n_{0}-1/2}{N_{00}%
}\bigg)^{2}+\bigg(\hat{k}_{1}^{\text{cut}}\frac{n_{1}}{N_{1}+1/2}%
\bigg)^{2}+1\bigg]^{-1}\;. \label{EQN005}%
\end{equation}
Since the left-hand sides of Eqs.\ (\ref{EQN004}) and (\ref{EQN005}) are
identical, we can equate their right-hand sides and eliminate $\lambda$,
\begin{align}
&  \sum_{n_{1}=-N_{1}}^{N_{1}}\bigg\{\frac{1}{N_{0c}}\sum_{n_{0}=1}^{N_{0c}%
}\bigg[\bigg(\hat{k}_{0}^{\text{cut}}\frac{n_{0}-1/2}{N_{0c}}\bigg)^{2}%
+\bigg(\hat{k}_{1}^{\text{cut}}\frac{n_{1}}{N_{1}+1/2}\bigg)^{2}%
\bigg]^{-1}\nonumber\label{EQN006}\\
&  \hspace{1cm}-\frac{1}{N_{00}}\sum_{n_{0}=1}^{N_{00}}\bigg[\bigg(\hat{k}%
_{0}^{\text{cut}}\frac{n_{0}-1/2}{N_{00}}\bigg)^{2}+\bigg(\hat{k}%
_{1}^{\text{cut}}\frac{n_{1}}{N_{1}+1/2}\bigg)^{2}+1\bigg]^{-1}\bigg\}=0\;.
\end{align}
For given $N_{0c}$, $N_{00}$, $N_{1}$, and $\hat{k}_{1}^{\text{cut}}$ one has
to solve this equation to obtain $\hat{k}_{0}^{\text{cut}}$. Then, $\lambda$
can be calculated using either Eq.\ (\ref{EQN004}) or (\ref{EQN005}).


\subsubsection{\label{SEC010}Optimizing $N_{0c}$}

The numerically obtained critical temperature $\hat{T}_{c} = \hat{k}%
_{0}^{\text{cut}} / 2 \pi N_{0c}$ should be insensitive with respect to
variations of $N_{0c}$, when keeping the other parameters fixed, in particular
$N_{00}$. The corresponding optimal $N_{0c}^{\text{opt}}$ is, therefore,
defined as the value of $N_{0c}$ which minimizes
\begin{align}
\label{EQN010}\bigg|\frac{\partial}{\partial N_{0c}} \hat{T}_{c}\bigg|
\end{align}
(since $N_{0c} \in\mathbb{N}$, the derivative $\partial/ \partial N_{0c}$ has
to be understood as a finite difference).

To study this optimization of $N_{0c}$ independently of any error due to the
finite spatial momentum cutoff $\hat{k}_{1}^{\text{cut}}$ and the finite
spatial extension $\hat{L}_{1}=2\pi(N_{1}+1/2)/\hat{k}_{1}^{\text{cut}}$, we
consider for a moment the limit $\hat{k}_{1}^{\text{cut}}\rightarrow\infty$
and $\hat{L}_{1}\rightarrow\infty$ (implying $N_{1}\rightarrow\infty$). In
this limit
\begin{equation}
\hat{k}_{1}^{\text{cut}}\frac{n_{1}}{N_{1}+1/2}\rightarrow\hat{k}_{1}%
\quad,\quad\frac{\hat{k}_{1}^{\text{cut}}}{N_{1}+1/2}\rightarrow d\hat{k}%
_{1}\quad,\quad\frac{N_{1}}{\hat{k}_{1}^{\text{cut}}}\sum_{n_{1}=-N_{1}%
}^{N_{1}}\frac{\hat{k}_{1}^{\text{cut}}}{N_{1}+1/2}\rightarrow\frac{N_{1}%
+1/2}{\hat{k}_{1}^{\text{cut}}}\int d\hat{k}_{1}\;.
\end{equation}
Inserting these relations into Eq.\ (\ref{EQN006}) and solving the integral
results in
\begin{equation}
\frac{1}{N_{0c}}\sum_{n_{0}=1}^{N_{0c}}\bigg(\hat{k}_{0}^{\text{cut}}%
\frac{n_{0}-1/2}{N_{0c}}\bigg)^{-1}-\frac{1}{N_{00}}\sum_{n_{0}=1}^{N_{00}%
}\bigg[\bigg(\hat{k}_{0}^{\text{cut}}\frac{n_{0}-1/2}{N_{00}}\bigg)^{2}%
+1\bigg]^{-1/2}=0\;. \label{EQN007}%
\end{equation}
As previously Eq.\ (\ref{EQN006}) this equation has to be solved to obtain
$\hat{k}_{0}^{\text{cut}}$, which now only depends on $N_{0c}$ and $N_{00}$.

\begin{figure}[tbh]
\centering
\includegraphics[width=0.485\textwidth]{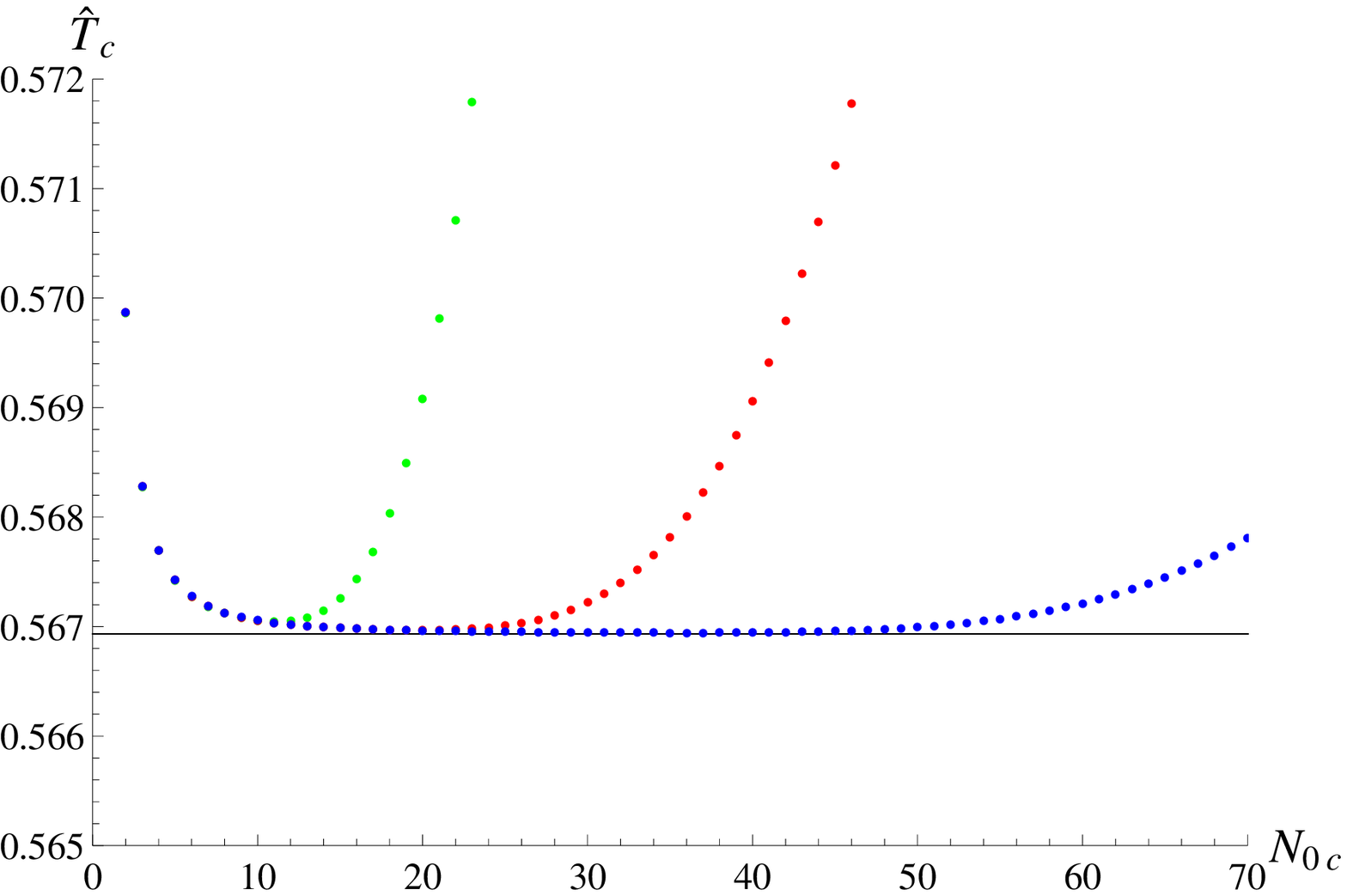}\hspace{0.4cm}%
\includegraphics[width=0.485\textwidth]{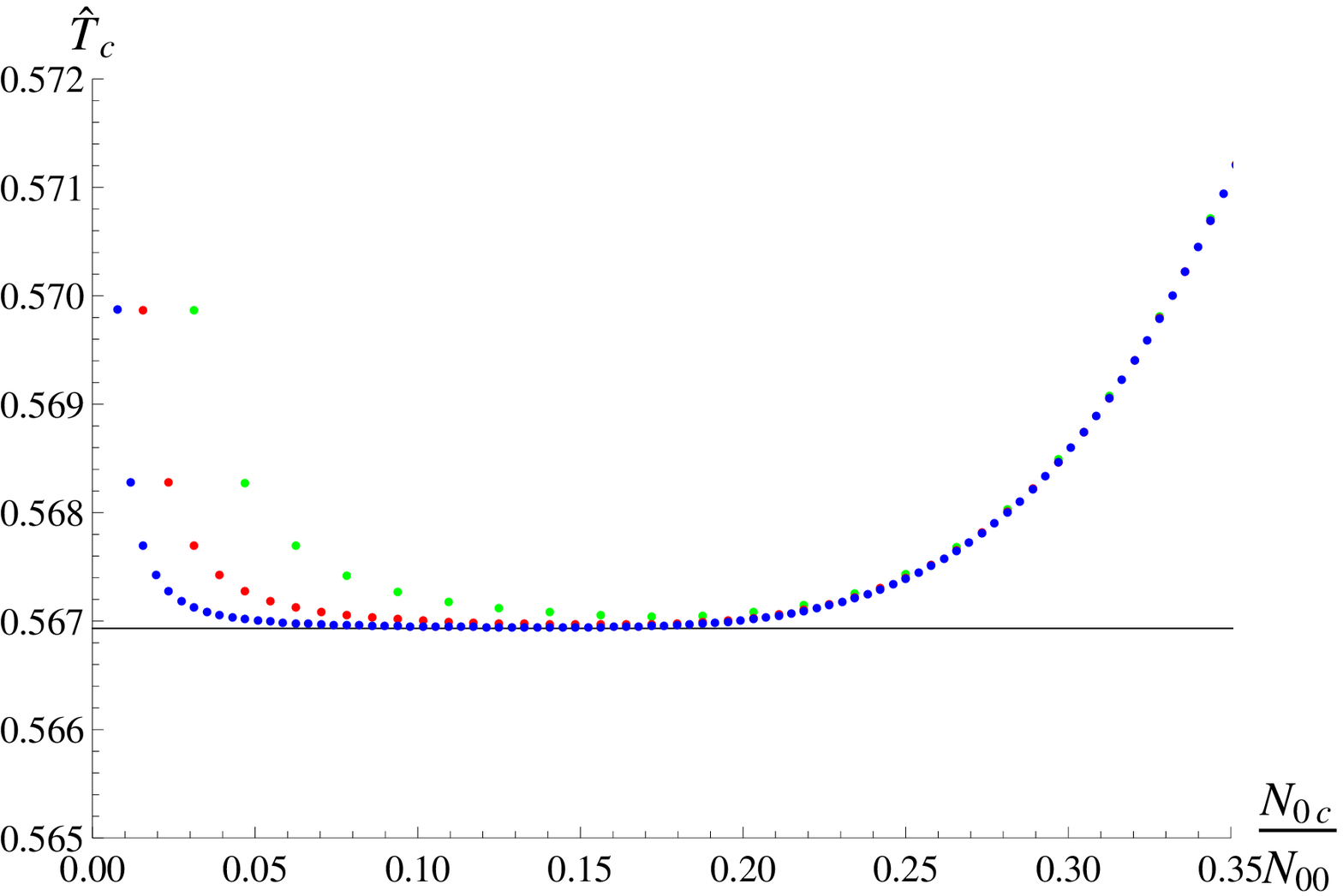}\caption{$\hat{T}_{c}$ as a
function of $N_{0c}$ (left panel) and $N_{0c} / N_{00}$ (right panel) for
$N_{00} = 64$ (green dots), $N_{00} = 128$ (red dots), and $N_{00} = 256$
(blue dots) obtained from Eq.\ (\ref{EQN007}) (the black lines indicate the
infinite-volume continuum result $\hat{T}_{c} = e^{C} / \pi$).}%
\label{Fig:N00_o_N0c}%
\end{figure}

In FIG.~\ref{Fig:N00_o_N0c} we study the corresponding $\hat{T}_{c} = \hat
{k}_{0}^{\text{cut}} / 2 \pi N_{0c}$ as a function of $N_{0c}$ (left panel)
and $N_{0c} / N_{00}$ (right panel) for $N_{00} \in\{ 64 , 128 , 256 \}$:

\begin{itemize}
\item For sufficiently large $N_{00}$ and a suitably chosen $N_{0c}$ the
resulting $\hat{T}_{c}$ should be close to the analytically known
infinite-volume continuum result $\hat{T}_{c} = e^{C} / \pi\approx0.566$
(where $C$ denotes Euler's constant) \cite{Wolff:1985av}. One can clearly see
that there are plateau-like regions, where this is the case.

\item For a small number of temporal modes $N_{0c}$ there are strong
deviations, because the temporal momentum cutoff is rather small, $\hat{k}%
_{0}^{\text{cut}} = 2 \pi N_{0c} \hat{T}_{c}$ (for small $N_{0c}$, curves
obtained with different $N_{00}$ fall on top of each other when plotted versus
$N_{0c} \propto\hat{k}_{0}^{\text{cut}}$).

\item For $N_{0c} / N_{00}
\raisebox{-0.5ex}{$\,\stackrel{>}{\scriptstyle\sim}\,$} 0.2$ there are also
strong deviations, because the temperature corresponding to $N_{00}$ temporal
modes, $\hat{T}_{0} = \hat{T}_{c} N_{0c} / N_{00}$, is a poor approximation of
zero temperature (for $N_{0c} / N_{00}
\raisebox{-0.5ex}{$\,\stackrel{>}{\scriptstyle\sim}\,$} 0.2$, curves obtained
with different $N_{00}$ fall on top of each other when plotted versus $N_{0c}
/ N_{00} = \hat{T}_{0} / \hat{T}_{c}$).
\end{itemize}

In other words, to obtain accurate results, $1 \ll N_{0c} \ll N_{00}$ has to
be fulfilled, which is only possible, if a sufficiently large number of
temporal modes $N_{00}$ are used. According to the definition (\ref{EQN010})
the optimal $N_{0c}$ for given $N_{00}$ is the minimum of the corresponding
curves in FIG.~\ref{Fig:N00_o_N0c}.

In TABLE~\ref{table:t_c_just_n00_n0c} one can see the accuracy $1-\hat{T}_{c}
/ (e^{C}/\pi)$ of the numerically obtained $\hat{T}_{c}$, for various values
of $N_{00}$. Note that even with a comparatively small number of $N_{00} = 32$
temporal modes, the error is less than $0.1 \%$. Also listed in
TABLE~\ref{table:t_c_just_n00_n0c} are $N_{0c}^{\text{opt}}$, the
corresponding temporal momentum cutoff $\hat{k}_{0}^{\text{cut,opt}} = 2 \pi
N_{0c}^{\text{opt}} \hat{T}_{c}$, and the corresponding temporal extension
$\hat{L}_{00}^{\text{opt}} = N_{00} / \hat{T}_{c} N_{0c}^{\text{opt}}$
approximating $T=0$. When increasing $N_{00}$, there is a similar increase in
the temporal momentum cutoff $\hat{k}_{0}^{\text{cut,opt}}$, but only a slight
increase in the temporal extension $\hat{L}_{00}^{\text{opt}}$. This is
typical for lattice calculations, where cutoff effects are only polynomially
suppressed, while finite-volume effects are exponentially suppressed.

\begin{table}[tbh]
\begin{center}
\begin{tabular}
[c]{c|c|c|c|c}%
$N_{00}$ & $1 - \hat{T}_{c}/(e^{C}/\pi)$ & $N_{0c}^{\text{opt}}$ & $\hat
{k}_{0}^{\text{cut,opt}}$ & $\hat{L}_{00}^{\text{opt}}$\\\hline\hline
$32$ & $-6.320 \cdot10^{-4}$ & $6$ & $2.139 \cdot10^{ 1}$ & $9.401$\\\hline
$64$ & $-1.981 \cdot10^{-4}$ & $11$ & $3.919 \cdot10^{ 1}$ & $10.26$\\\hline
$128$ & $-6.158 \cdot10^{-5}$ & $20$ & $7.125 \cdot10^{ 1}$ & $11.29$%
\\\hline\hline
$256$ & $-1.881 \cdot10^{-5}$ & $36$ & $1.282 \cdot10^{ 2}$ & $12.54$%
\\\hline\hline
$512$ & $-5.650 \cdot10^{-6}$ & $66$ & $2.350 \cdot10^{ 2}$ & $13.68$\\\hline
$1024$ & $-1.672 \cdot10^{-6}$ & $120$ & $4.275 \cdot10^{ 2}$ & $15.05$%
\\\hline
$2048$ & $-4.886 \cdot10^{-7}$ & $222$ & $7.908 \cdot10^{ 2}$ & $16.27$%
\\\hline
$4096$ & $-1.413 \cdot10^{-7}$ & $412$ & $1.468 \cdot10^{ 3}$ & $17.53$%
\end{tabular}
\end{center}
\caption{Accuracy of the numerically obtained $\hat{T}_{c}$ for various values
of $N_{00}$.}%
\label{table:t_c_just_n00_n0c}%
\end{table}


\subsubsection{\label{SEC650}Optimizing $\hat{k}_{1}^{\text{cut}}$}

In the following we investigate the error associated with a finite spatial
momentum cutoff $\hat{k}_{1}^{\text{cut}}$, i.e., instead of
Eq.\ (\ref{EQN007}) we return to Eq.\ (\ref{EQN006}) and solve this equation
to obtain $\hat{k}_{0}^{\text{cut}}$ for given $N_{0c}$, $N_{00}$, $N_{1}$,
and $\hat{k}_{1}^{\text{cut}}$. Similarly to Sec.~\ref{SEC010} we define the
optimal $\hat{k}_{1}^{\text{cut},\text{opt}}$ as the value of $\hat{k}%
_{1}^{\text{cut}}$ which minimizes
\begin{equation}
\bigg|\frac{\partial}{\partial\hat{k}_{1}^{\text{cut}}}\hat{T}_{c}%
\bigg| \label{EQN011}%
\end{equation}
for given $N_{0c}$, $N_{00}$, and $N_{1}$.

We choose $N_{00} = 256$, the corresponding optimal $N_{0c} = 36$
(cf.\ TABLE~\ref{table:t_c_just_n00_n0c}) and various numbers of spatial modes
$N_{1}$. In FIG.~\ref{Fig:fhat_vs_N} we study $\hat{T}_{c} = \hat{k}%
_{0}^{\text{cut}} / 2 \pi N_{0c}$ as a function of $\hat{k}_{1}^{\text{cut}}$
(left panel) and $\hat{k}_{1}^{\text{cut}}/(N_{1} + 1/2)$ (right panel)
obtained from Eq.\ (\ref{EQN006}) for $N_{1} \in\{ 64 , 128 , 256 \}$:

\begin{figure}[tbh]
\centering
\includegraphics[width=0.485\textwidth]{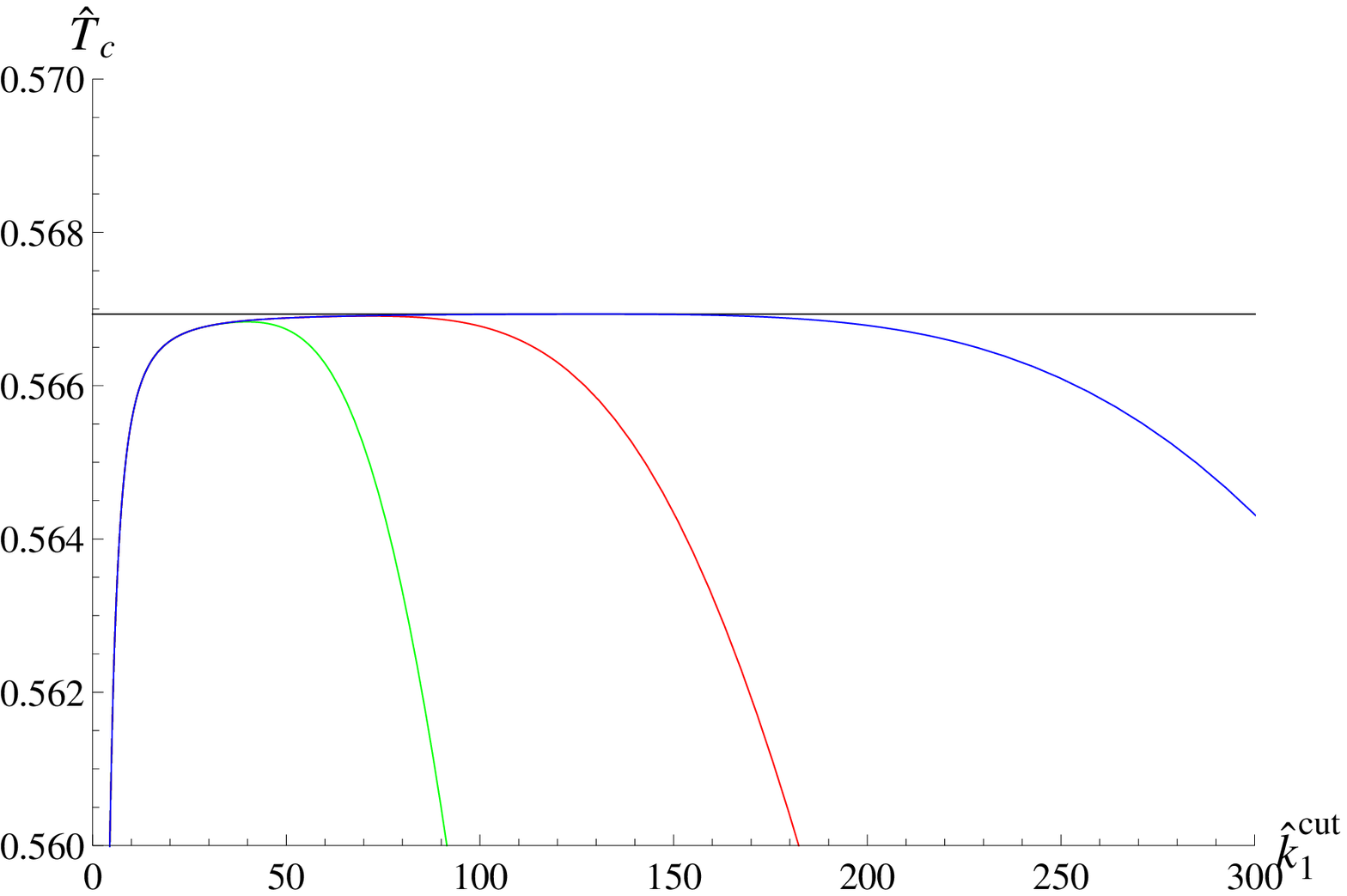}\hspace{0.4cm}%
\includegraphics[width=0.485\textwidth]{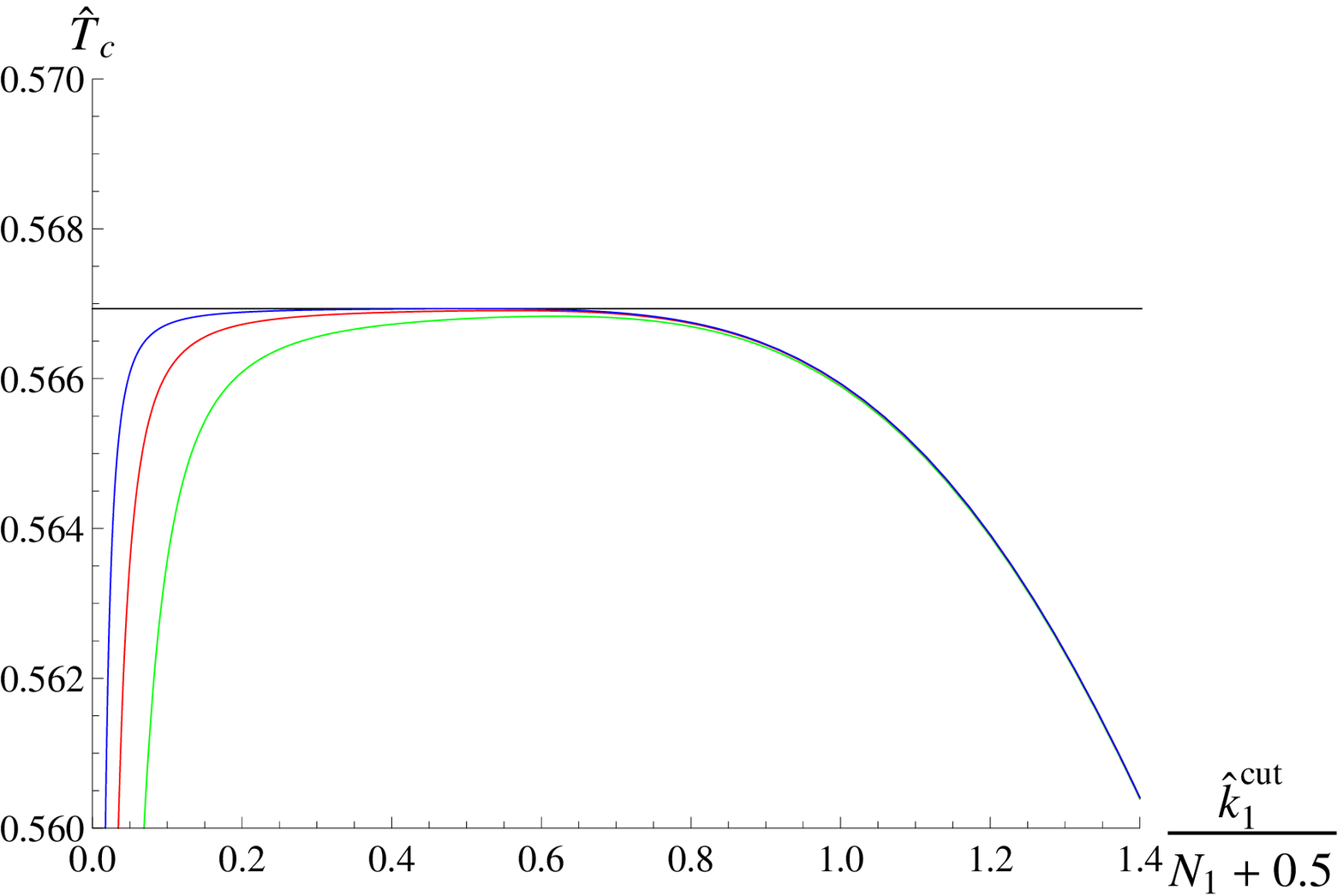}\caption{$\hat
{T}_{c}$ as a function of $\hat{k}_{1}^{\text{cut}}$ (left panel) and $\hat
{k}_{1}^{\text{cut}} / (N_{1} + 1/2)$ (right panel) for $N_{1} = 64$ (green
curve), $N_{1} = 128$ (red curve), and $N_{1} = 256$ (blue curve) obtained
from (\ref{EQN006}) with $N_{00} = 256$ and $N_{0c} = 36$ (the black lines
indicate the infinite-volume continuum result $\hat{T}_{c} = e^{C} / \pi$).}%
\label{Fig:fhat_vs_N}%
\end{figure}

\begin{itemize}
\item For sufficiently large $N_{1}$ and a suitably chosen $\hat{k}%
_{1}^{\text{cut}}$ the resulting $\hat{T}_{c}$ should be close to the
analytically known infinite-volume continuum result $\hat{T}_{c} = e^{C} /
\pi\approx0.566$. Again one can observe plateau-like regions, where this is
the case.

\item For a small spatial momentum cutoff $\hat{k}_{1}^{\text{cut}}$ there are
strong deviations (for small $\hat{k}_{1}^{\text{cut}}$, curves obtained with
different $N_{1}$ fall on top of each other, when plotted versus $\hat{k}%
_{1}^{\text{cut}}$).

\item For $\hat{k}_{1}^{\text{cut}} / (N_{1} + 1/2)
\raisebox{-0.5ex}{$\,\stackrel{>}{\scriptstyle\sim}\,$} 1.0$ there are also
strong deviations, because the extent of the periodic spatial dimension
$\hat{L}_{1} = 2 \pi(N_{1} + 1/2) / \hat{k}_{1}^{\text{cut}}$ is quite small
and, therefore, a poor approximation for infinitely extended space (for
$\hat{k}_{1}^{\text{cut}} / (N_{1} + 1/2)
\raisebox{-0.5ex}{$\,\stackrel{>}{\scriptstyle\sim}\,$} 1.0$, curves obtained
with different $N_{1}$ fall on top of each other, when plotted versus $\hat
{k}_{1}^{\text{cut}} / (N_{1} + 1/2) \propto1 / \hat{L}_{1}$).
\end{itemize}

In other words, to obtain accurate results, $1 \ll\hat{k}_{1}^{\text{cut}} \ll
N_{1}$ has to be fulfilled, which is only possible, if a sufficiently large
number of spatial modes $N_{1}$ is used. According to the definition
(\ref{EQN011}) the optimal $\hat{k}_{1}^{\text{cut}}$ for given $N_{1}$ is the
maximum of the corresponding curves in FIG.~\ref{Fig:fhat_vs_N}.

In TABLE~\ref{table:t_c_n} one can see the accuracy $1-\hat{T}_{c} /
(e^{C}/\pi)$ of the numerically obtained $\hat{T}_{c}$ for various values of
$N_{1}$, again for $N_{00} = 256$ and the corresponding optimal $N_{0c} = 36$.
Note that errors due to the finite-mode regularization in temporal direction
and in spatial direction have opposite sign (cf.\ FIG.~\ref{Fig:N00_o_N0c} and
FIG.~\ref{Fig:fhat_vs_N}). Consequently, one obtains the most accurate result
for $\hat{T}_{c}$ not for $N_{1} \rightarrow\infty$, but when both errors
almost cancel each other. This is the case for $N_{1} = 249$, i.e., for
$N_{1}$ similar to $N_{00}$ (for a more detailed discussion
cf.\ Sec.~\ref{SEC011}). Also listed in TABLE~\ref{table:t_c_n} are the
spatial momentum cutoff $\hat{k}_{1}^{\text{cut,opt}}$ and the corresponding
spatial extension $\hat{L}_{1}^{\text{opt}} = 2 \pi(N_{1} + 1/2) / \hat{k}%
_{1}^{\text{cut,opt}}$. Again, when increasing $N_{1}$, there is a similar
increase in the spatial momentum cutoff $\hat{k}_{1}^{\text{cut,opt}}$, but
only a slight increase in the spatial extension $\hat{L}_{1}^{\text{opt}}$
approximating infinite volume. As already mentioned this is typical for
lattice calculations, where cutoff effects are only polynomially suppressed,
while finite-volume effects are exponentially suppressed.

\begin{table}[tbh]
\begin{center}
\begin{tabular}
[c]{c|c|c|c}%
$N_{1}$ & $1 - \hat{T}_{c}/(e^{C}/\pi)$ & $\hat{k}_{1}^{\text{cut,opt}}$ &
$\hat{L}_{1}^{\text{opt}}$\\\hline\hline
$32$ & $+5.789 \cdot10^{-4}$ & $2.273 \cdot10^{1}$ & $\phantom{0}8.99$\\\hline
$64$ & $+1.766 \cdot10^{-4}$ & $3.991 \cdot10^{1}$ & $10.15$\\\hline
$128$ & $+4.324 \cdot10^{-5}$ & $7.138 \cdot10^{1}$ & $11.31$\\\hline\hline
$249$ & $-2.720 \cdot10^{-8}$ & $1.261 \cdot10^{2}$ & $12.43$\\\hline\hline
$256$ & $-1.003 \cdot10^{-6}$ & $1.291 \cdot10^{2}$ & $12.48$\\\hline
$512$ & $-1.460 \cdot10^{-5}$ & $2.329 \cdot10^{2}$ & $13.83$\\\hline
$1024$ & $-1.798 \cdot10^{-5}$ & $4.179 \cdot10^{2}$ & $15.41$\\\hline
$2048$ & $-1.871 \cdot10^{-5}$ & $7.616 \cdot10^{2}$ & $16.90$\\\hline
$4096$ & $-1.882 \cdot10^{-5}$ & $1.403 \cdot10^{3}$ & $18.35$%
\end{tabular}
\end{center}
\caption{Accuracy of the numerically obtained $\hat{T}_{c}$ for $N_{00} =
256$, $N_{0c} = 36$, and various values of $N_{1}$.}%
\label{table:t_c_n}%
\end{table}


\subsubsection{\label{SEC011}Optimizing the ratio of $N_{00}$ and $N_{1}$}

As already mentioned the maximum number of modes $\propto N_{00} N_{1}$ is
limited by the available computer resources. In the previous subsection it has
been observed that for $N_{00} = 256$ and $N_{1} = 249$ the errors in $\hat
{T}_{c}$ due to the finite-mode regularization almost cancel. Also for other
choices of $N_{00}$ such a nearly perfect cancellation is present for $N_{1}
\approx N_{00}$, as collected in TABLE~\ref{TAB001}. Moreover, note that the
temporal momentum cutoff $\hat{k}_{0}^{\text{cut,opt}}$ and the spatial
momentum cutoff $\hat{k}_{1}^{\text{cut,opt}}$ are close to each other.

\begin{figure}[tbh]
\centering
\includegraphics[width=0.50\textwidth]{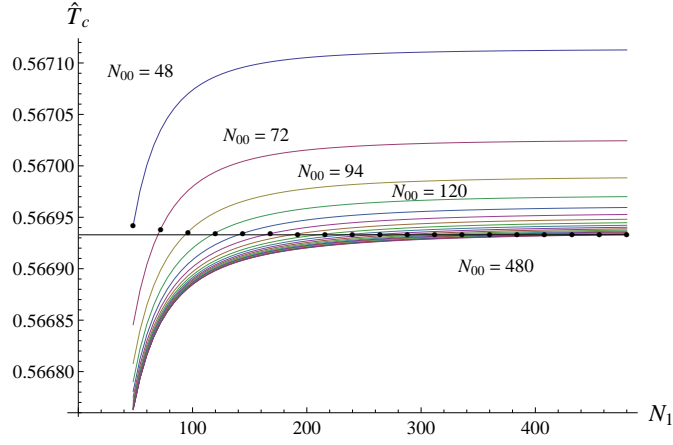}\caption{$\hat{T}_{c}$
as a function of $N_{1}$ for several $N_{00}$ (filled black circles indicate
symmetric choices $N_{00} = N_{1}$, the black line is the infinite-volume
continuum result $\hat{T}_{c} = e^{C} / \pi$).}%
\label{Fig:N_i_N_0_full}%
\end{figure}

Similarly, in FIG.~\ref{Fig:N_i_N_0_full} we compare the numerically obtained
$\hat{T}_{c}$ for various $N_{00}$ and $N_{1}$ with the infinite-volume
continuum result $\hat{T}_{c} = e^{C} / \pi$. The figure suggests choosing
$N_{00} = N_{1}$ as a simple rule, which obviously leads to very accurate
numerical results (the black filled circles in FIG.~\ref{Fig:N_i_N_0_full}).
Unless mentioned otherwise, we will use $N_{00} = N_{1}$ in the following. Of
course, such a cancellation of errors might not occur for quantities other
than $\hat{T}_{c}$. Moreover, when considering also the possibility of an
inhomogeneous condensate, as we will do in Sec.~\ref{SEC599}, it could be
necessary to have a finer resolution or a larger extent of the spatial
dimension, which might require a rather large $N_{1} \gg N_{00}$.

\begin{table}[tbh]
\begin{center}
\begin{tabular}
[c]{c|c|c|c|c}%
$N_{00}$ & $N_{1}$ & $N_{0c}^{\text{opt}}$ & $\hat{k}_{0}^{\text{cut,opt}}$ &
$\hat{k}_{1}^{\text{cut,opt}}$\\\hline\hline
$32$ & $30$ & $6$ & $2.137 \cdot10^{1}$ & $2.190 \cdot10^{1}$\\\hline
$64$ & $62$ & $11$ & $3.919 \cdot10^{1}$ & $3.932 \cdot10^{1}$\\\hline
$128$ & $125$ & $20$ & $7.124 \cdot10^{1}$ & $7.034 \cdot10^{1}$\\\hline\hline
$256$ & $249$ & $36$ & $1.282 \cdot10^{2}$ & $1.261 \cdot10^{2}$\\\hline\hline
$512$ & $497$ & $66$ & $2.351 \cdot10^{2}$ & $2.287 \cdot10^{2}$\\\hline
$1024$ & $994$ & $121$ & $4.310 \cdot10^{2}$ & $4.188 \cdot10^{2}$\\\hline
$2048$ & $1985$ & $223$ & $7.844 \cdot10^{2}$ & $7.710 \cdot10^{2}$\\\hline
$4096$ & $3966$ & $414$ & $1.475 \cdot10^{3}$ & $1.428 \cdot10^{3}$%
\end{tabular}
\end{center}
\caption{Pairs $N_{00}$ and $N_{1}$, where the errors due to the finite-mode
regularization for $\hat{T}_{c}$ almost cancel, and the corresponding
$N_{0c}^{\text{opt}}$, $\hat{k}_{0}^{\text{cut,opt}}$, and $\hat{k}%
_{1}^{\text{cut,opt}}$.}%
\label{TAB001}%
\end{table}

\subsubsection{Summary}

Based on these investigations we propose and adopt the following strategy to
determine the parameters $N_{0c}$, $N_{1}$, $\hat{k}_{0}^{\text{cut}}$,
$\hat{k}_{1}^{\text{cut}}$, and $\lambda$:

\begin{itemize}
\item[(1)] Use $N_{00} = N_{1}$ as large as possible (limited by the available
computer resources).

\item[(2)] Determine the corresponding optimal $N_{0c}$ from a computation in
the limit $N_{1} \rightarrow\infty$ and $\hat{k}_{1}^{\text{cut}}
\rightarrow\infty$ as described in Sec.~\ref{SEC010} (cf.\ also
TABLE~\ref{table:t_c_just_n00_n0c}, third column). This computation also
provides a value for $\hat{k}_{0}^{\text{cut}}$
(TABLE~\ref{table:t_c_just_n00_n0c}, fifth column), which is a good
approximation for $\hat{k}_{0}^{\text{cut}}$ at finite, but large $N_{1}$ and
$\hat{k}_{1}^{\text{cut}}$. Assign this value to $\hat{k}_{1}^{\text{cut}}$
(cf.\ Sec.~\ref{SEC011}).

\item[(3)] Solve Eq.\ (\ref{EQN006}) to determine $\hat{k}_{0}^{\text{cut}}$
(now for finite $N_{1}$ and $\hat{k}_{1}^{\text{cut}}$) for the previously
chosen $N_{00}$ and $N_{1}$ [step~(1)] and $N_{0c}$ and $\hat{k}%
_{1}^{\text{cut}}$ [step~(2)].

\item[(4)] Determine $\lambda$ via Eq.\ (\ref{EQN004}) [or equivalently via
Eq.\ (\ref{EQN005})].
\end{itemize}

For all further computations, e.g.\ when computing the phase diagram for a
homogeneous condensate $\sigma= \text{constant}$ or an inhomogeneous
condensate $\sigma= \sigma(x_{1})$, the parameters $N_{1}$, $\hat{k}%
_{0}^{\text{cut}}$, $\hat{k}_{1}^{\text{cut}}$, and $\lambda$ are not changed
anymore. The temperature $\hat{T} = 1 / \hat{L}_{0} = \hat{k}_{0}^{\text{cut}}
/ 2 \pi N_{0}$ can be adjusted by using different numbers of temporal modes
$N_{0}$. $N_{0c}$ corresponds to the critical temperature $\hat{T}_{c}$.


\subsection{\label{SEC599}Computation of the phase diagram for a homogeneous
condensate $\sigma= \text{constant}$}

To determine the phase diagram for a homogeneous condensate, i.e., $\sigma$ as
a function of the chemical potential $\mu$ and the temperature $T$, one
proceeds as in the previous subsection. From
\begin{equation}
0=\frac{d}{d\hat{\sigma}}\frac{S_{E,\text{eff}}(\hat{\sigma})}{N}%
\end{equation}
one can derive the generalization of Eq.\ (\ref{EQN003}) for arbitrary
$\mu\geq0$,
\begin{equation}
0=\frac{2\pi^{2}N_{0}(N_{1}+1/2)}{\lambda\hat{k}_{0}^{\text{cut}}\hat{k}%
_{1}^{\text{cut}}}-2\sum_{n_{0}=1}^{N_{0}}\sum_{n_{1}=-N_{1}}^{N_{1}}%
\frac{\bigg(\hat{k}_{0}^{\text{cut}}\frac{n_{0}-1/2}{N_{0}}\bigg)^{2}%
+\bigg(\hat{k}_{1}^{\text{cut}}\frac{n_{1}}{N_{1}+1/2}\bigg)^{2}+\hat{\sigma
}^{2}-\hat{\mu}^{2}}{\bigg[\bigg(\hat{k}_{0}^{\text{cut}}\frac{n_{0}%
-1/2}{N_{0}}\bigg)^{2}+\bigg(\hat{k}_{1}^{\text{cut}}\frac{n_{1}}{N_{1}%
+1/2}\bigg)^{2}+\hat{\sigma}^{2}-\hat{\mu}^{2}\bigg]^{2}+\bigg(2\hat{\mu}%
\hat{k}_{0}^{\text{cut}}\frac{n_{0}-1/2}{N_{0}}\bigg)^{2}}\;. \label{EQN942}%
\end{equation}
If this equation has a solution $\hat{\sigma}^{2}=A>0$ for given $(\hat{\mu
},\hat{T})$, and if $S_{E,\text{eff}}(\sqrt{A})<S_{E,\text{eff}}(0)$, this
solution is the value of the chiral condensate, i.e., $\hat{\sigma}=\pm
\sqrt{A}$, and $(\hat{\mu},\hat{T})$ is inside the chirally broken phase. If
there is no such solution, or if $S_{E,\text{eff}}(\sqrt{A})\geq
S_{E,\text{eff}}(0)$, then $\hat{\sigma}=0$ and $(\hat{\mu},\hat{T})$ is a
point inside the chirally symmetric phase or on the phase boundary.

\begin{figure}[tbh]
\centering
\includegraphics[width=0.50\textwidth]{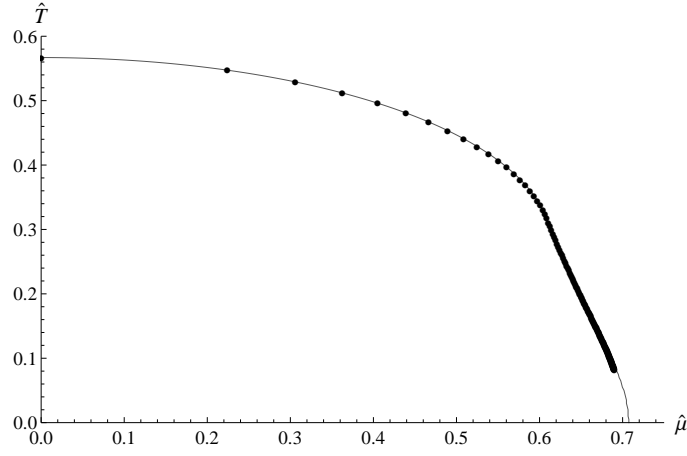}\caption{Phase diagram of
the GN model for a homogeneous condensate $\sigma= \text{constant}$ for
$N_{00} = N_{1} = 192$ [the light gray curve is the infinite-volume continuum
result \cite{Wolff:1985av}].}%
\label{FIG008}%
\end{figure}

The phase diagram for a homogeneous condensate obtained with $N_{00} = N_{1} =
192$ (with corresponding $N_{0c} = 28$, $\hat{k}_{0}^{\text{cut}} = 9.974
\cdot10^{1}$, $\hat{k}_{1}^{\text{cut}} = 1.011 \cdot10^{2}$ and $\lambda=
0.3328$) is shown in FIG.~\ref{FIG008}. It is in excellent agreement with the
infinite-volume continuum result \cite{Wolff:1985av}.


\subsection{\label{SEC470}Finite-mode regularization, spatially inhomogeneous
condensate $\sigma= \sigma(x_{1})$}

To study the possibility of a spatially inhomogeneous condensate, $\hat
{\sigma}$ is written as a superposition of a finite number of plane waves,
\begin{equation}
\hat{\sigma}(\hat{x}_{1})=\frac{\sigma(x_{1})}{\sigma_{0}}=\sum_{m=-M}%
^{M}c_{m}\frac{e^{-\imath\hat{p}\hat{x}_{1}}}{\sqrt{\hat{L}_{1}}}\quad
,\quad\hat{p}=\frac{2\pi}{\hat{L}_{1}}m\quad,\quad c_{+m}=(c_{-m})^{\ast},
\label{EQN751}%
\end{equation}
as done in Eqs.\ (\ref{EQN050}) and (\ref{EQN051}) for the fermionic fields
$\hat{\psi}_{j}$. The resolution of $\hat{\sigma}$ should be coarser than the
resolution of $\hat{\psi}_{j}$, i.e., $M\ll N_{1}$, to obtain stable and
meaningful numerical results [cf.\ Ref.\ \cite{Wagner:2007he} for a detailed discussion].

In the case of a spatially inhomogeneous condensate plane waves are no longer
eigenfunctions of the Dirac operator $Q$. Consequently, $\ln[\det(\hat
{Q}^{\dagger}\hat{Q})]$ cannot be expressed as a product over modes as done in
Eq.\ (\ref{EQN766}). One has to represent $\hat{Q}^{\dagger}\hat{Q}$ as a
matrix, where the rows and columns correspond to the plane-wave basis
functions of the fermionic fields $e^{\mp\imath(\hat{k}_{0}\hat{x}_{0}+\hat
{k}_{1}\hat{x}_{1})}/\sqrt{\hat{L}_{0}\hat{L}_{1}}$ [cf.\ Eqs.\ (\ref{EQN050})
and (\ref{EQN051})],
\begin{align}
&  \langle\hat{k}_{0},\hat{k}_{1}|\hat{Q}^{\dagger}\hat{Q}|\hat{k}_{0}%
^{\prime},\hat{k}_{1}^{\prime}\rangle=\nonumber\\
&  \hspace{0.5cm}=\frac{1}{\hat{L}_{0}\hat{L}_{1}}\int_{0}^{\hat{L}_{0}}%
d\hat{x}_{0}\,\int_{0}^{\hat{L}_{1}}d\hat{x}_{1}\,e^{+\imath(\hat{k}_{0}%
\hat{x}_{0}+\hat{k}_{1}\hat{x}_{1})}\bigg(-\gamma_{\mu}\hat{\partial}_{\mu
}+\gamma_{0}\hat{\mu}+\sum_{m=-M}^{+M}c_{m}\frac{e^{-\imath\hat{p}\hat{x}_{1}%
}}{\sqrt{\hat{L}_{1}}}\bigg)\nonumber\\
&  \hspace{1cm}\times\bigg(+\gamma_{\mu}\hat{\partial}_{\mu}+\gamma_{0}%
\hat{\mu}+\sum_{m^{\prime}=-M}^{+M}c_{m^{\prime}}\frac{e^{-\imath\hat
{p}^{\prime}\hat{x}_{1}}}{\sqrt{\hat{L}_{1}}}\bigg)e^{-\imath(\hat{k}%
_{0}^{\prime}\hat{x}_{0}+\hat{k}_{1}^{\prime}\hat{x}_{1})}\nonumber\\
&  \hspace{0.5cm}=\frac{\delta_{\hat{k}_{0},\hat{k}_{0}^{\prime}}}{\hat{L}%
_{1}}\int_{0}^{\hat{L}_{1}}d\hat{x}_{1}\,e^{+\imath\hat{k}_{1}\hat{x}_{1}%
}\bigg(+\imath\gamma_{0}\hat{k}_{0}+\imath\gamma_{1}\hat{k}_{1}+\gamma_{0}%
\hat{\mu}+\sum_{m=-M}^{+M}c_{m}\frac{e^{-\imath\hat{p}\hat{x}_{1}}}{\sqrt
{\hat{L}_{1}}}\bigg)\nonumber\\
&  \hspace{1cm}\times\bigg(-\imath\gamma_{0}\hat{k}_{0}^{\prime}-\imath
\gamma_{1}\hat{k}_{1}^{\prime}+\gamma_{0}\hat{\mu}+\sum_{m^{\prime}=-M}%
^{+M}c_{m^{\prime}}\frac{e^{-\imath\hat{p}^{\prime}\hat{x}_{1}}}{\sqrt{\hat
{L}_{1}}}\bigg)e^{-\imath\hat{k}_{1}^{\prime}\hat{x}_{1}}%
\end{align}
with $\hat{p}=2\pi m/\hat{L}_{1}$ and $\hat{p}^{\prime}=2\pi m^{\prime}%
/\hat{L}_{1}$. These matrix elements can be calculated analytically. Note that
this matrix representation of $\det(\hat{Q}^{\dagger}\hat{Q})$ has a
block-diagonal structure with $2N_{0}$ blocks of size $2(2N_{1}+1)$ (the
blocks are labeled by $\hat{k}_{0}=\hat{k}_{0}^{\prime}$; the rows and columns
of each block correspond to the spatial momenta $\hat{k}_{1}$ and $\hat{k}%
_{1}^{\prime}$ and the two spin components). $\ln[\det(\hat{Q}^{\dagger}%
\hat{Q})]$ is then the sum over $\ln[\det(\ldots)]$ of the blocks, where each
term of that sum can be evaluated numerically e.g.\ by means of an LU decomposition.


\subsection{Computation of the phase diagram for an inhomogeneous condensate
$\sigma= \sigma(x_{1})$}

When allowing for a spatially inhomogeneous condensate $\sigma= \sigma(x_{1}%
)$, there are three phases:

\begin{itemize}
\item[(I)] For small chemical potential $\hat{\mu}$ and low temperature
$\hat{T}$ chiral symmetry is broken by a homogeneous condensate $\hat{\sigma}
= \text{constant} \neq0$ [corresponding to $c_{0} \neq0$, $c_{m} = 0$ for $m
\neq0$ in Eq.\ (\ref{EQN751})].

\item[(II)] For high temperature $\hat{T}$ chiral symmetry is intact and
$\hat{\sigma} = 0$ [corresponding to $c_{m} = 0$ for all $m$ in
Eq.\ (\ref{EQN751})].

\item[(III)] For large chemical potential $\hat{\mu}$ and low temperature
$\hat{T}$ there is a spatially inhomogeneous condensate $\hat{\sigma} =
\hat{\sigma}(\hat{x}_{1})$ [corresponding to $c_{m} \neq0$ for at least one $m
\neq0$ in Eq.\ (\ref{EQN751})].
\end{itemize}

\begin{figure}[tbh]
\centering
\includegraphics[width=0.50\textwidth]{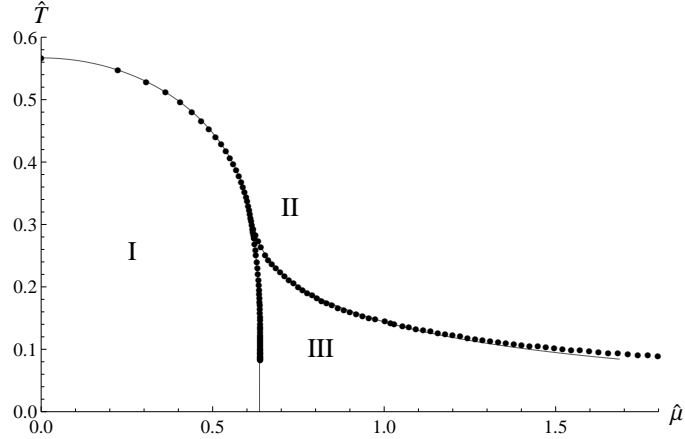}\caption{Phase diagram of
the GN model for a spatially inhomogeneous condensate $\sigma= \sigma(x_{1})$
for $N_{00} = N_{1} = 192$ and $M = 10$ [the light gray curve is the
infinite-volume continuum result \cite{Thies:2003kk,Schnetz:2004vr}].}%
\label{Fig:GN_pd_192}%
\end{figure}

The phase diagram for a spatially inhomogeneous condensate obtained with
$N_{00} = N_{1} = 192$ and $M = 10$ (with corresponding $N_{0c} = 28$,
$\hat{k}_{0}^{\text{cut}} = 9.974 \cdot10^{1}$, $\hat{k}_{1}^{\text{cut}} =
1.011 \cdot10^{2}$ and $\lambda= 0.3328$) is shown in FIG.~\ref{Fig:GN_pd_192}%
. It is in excellent agreement with the infinite-volume continuum result
\cite{Thies:2003kk,Schnetz:2004vr}.

The numerical determination of the phase boundaries is discussed in detail in
Ref.\ \cite{Wagner:2007he} and, therefore, only summarized briefly in the following.

\begin{itemize}
\item \textbf{Phase boundary I-II:} \newline The phase boundary between
$\hat{\sigma}= \text{constant} \neq0$ (phase I) and $\hat{\sigma}= 0$ (phase
II) can be determined as explained in Sec.~\ref{SEC599} for the phase diagram
for a homogeneous condensate.

\item \textbf{Phase boundary I-III:} \newline To determine the phase boundary
between $\hat{\sigma}= \text{constant} \neq0$ (phase I) and the inhomogeneous
crystal phase (phase III), one again has to find the minimum of
$S_{E,\text{eff}} / N$ with respect to $\hat{\sigma}$ as a function of
$(\hat{\mu},\hat{T})$. This time, however, $\hat{\sigma}$ is not a constant,
but a superposition of plane waves [cf.\ Eq.\ (\ref{EQN751})]. The
minimization has to be done with respect to the coefficients $c_{n}$.

\item \textbf{Phase boundary II-III:} \newline To determine the phase boundary
between $\hat{\sigma}= 0$ (phase II) and the inhomogeneous crystal phase
(phase III), one can in principle proceed as for the phase boundary I-III.
Note, however, that inside the crystal phase in the vicinity of the phase
boundary I-III the constant $\hat{\sigma}$ of the phase diagram for a
homogeneous condensate is a local minimum (the corresponding phase transition
is of first order), while in the vicinity of the phase boundary II-III it is a
saddle point (the corresponding phase transition is of second order).
Therefore, a computationally simpler and cheaper way to determine the phase
boundary II-III is to study the smallest eigenvalue of the Hessian matrix
\begin{align}
H_{m m^{\prime}} = \frac{\partial}{\partial C_{m}} \frac{\partial}{\partial
C_{m^{\prime}}} \frac{S_{E,\text{eff}}(\hat{\sigma})}{N}\bigg|_{\hat{\sigma
}=0} \quad, \quad C_{2 m} = \text{Re}(c_{m}) \quad, \quad C_{2 m + 1} =
\text{Im}(c_{m}) \; ,
\end{align}
with $0 < m , m^{\prime}\leq M$. A negative eigenvalue amounts to a direction
of negative curvature and, therefore, indicates the existence of an
inhomogeneous condensate.
\end{itemize}

Since the finite-mode approach allows to determine the condensate $\sigma$ at
given temperature $T$ for arbitrary chemical potential $\mu$, it is
straightforward to study and reproduce the order of the transition along the
phase boundaries I-II, I-III, and II-III.


\section{\label{SEC_chiGN}Phase diagram of the $1+1$ dimensional $\chi$GN
model}

We proceed in the same way as explained in detail in the previous section for
the GN model. After introducing two real scalar fields $\sigma$ and $\eta$,
the partition function of the $\chi$GN model can be written as
\begin{equation}
Z=\int D\sigma\,D\eta\,e^{-S_{E,\text{eff}}[\sigma,\eta]}\quad,\quad
S_{E,\text{eff}}[\sigma,\eta]=N\int d^{2}x\,\bigg\{\frac{1}{2\lambda
}\Big(\sigma^{2}+\eta^{2}\Big)-\frac{1}{2}\ln[\det(Q^{\dagger}Q)]\bigg\}\;,
\label{EQN156}%
\end{equation}
where
\[
Q=\gamma_{\mu}\partial_{\mu}+\gamma_{0}\mu+\sigma+\eta\imath\gamma_{5}.
\]
We then apply the finite-mode regularization, i.e., in analogy to
Eq.\ (\ref{EQN751}) the scalar fields $\sigma$ and $\eta$ (the condensates)
are represented as a sum over a finite number of modes,
\begin{align}
&  \hat{\sigma}(\hat{x}_{1})=\frac{\sigma(x_{1})}{\Sigma_{0}}=\sum_{m=-M}%
^{M}c_{m}\frac{e^{-\imath\hat{p}\hat{x}_{1}}}{\sqrt{\hat{L}_{1}}}\quad,\quad
c_{+m}=(c_{-m})^{\ast}\quad,\label{EQN607}\\
&  \hat{\eta}(\hat{x}_{1})=\frac{\eta(x_{1})}{\Sigma_{0}}=\sum_{m=-M}^{M}%
d_{m}\frac{e^{-\imath\hat{p}\hat{x}_{1}}}{\sqrt{\hat{L}_{1}}}\quad,\quad
d_{+m}=(d_{-m})^{\ast}\quad,\quad\hat{p}=\frac{2\pi}{\hat{L}_{1}}m\;.
\label{EQN608}%
\end{align}
Similary, $Q^{\dagger}Q$ is written as a matrix, where the rows and columns
correspond to plane-wave basis functions $e^{\mp\imath(\hat{k}_{0}\hat{x}%
_{0}+\hat{k}_{1}\hat{x}_{1})}/\sqrt{\hat{L}_{0}\hat{L}_{1}}$,
[cf.\ Sec.~\ref{SEC470} and Eqs.\ (\ref{EQN050}) and (\ref{EQN051}) for
details]. Since $S_{E,\text{eff}}$ is invariant under the transformation
$(\sigma,\eta)\rightarrow R(\sigma,\eta)$ with $R\in\text{O(2)}$, dimensionful
quantities are expressed in units of $\Sigma_{0}$, where
\begin{equation}
\Sigma_{0}=\Sigma\Big|_{T=0,\mu=0}\quad,\quad\Sigma=\Big(\sigma^{2}+\eta
^{2}\Big)^{1/2}\;, \label{EQN799}%
\end{equation}
and denoted by a hat $\hat{\phantom{x}}$.

We have studied the phase diagram of the $\chi$GN model using $M = 10$ modes
for the condensates and $N_{00} = N_{1} = 96$ modes for the fermionic determinant.

For temperatures $\hat{T}>\hat{T}_{c}=e^{C}/\pi$ and arbitrary chemical
potential $\mu$ chiral symmetry is restored, i.e., the effective action
(\ref{EQN156}) is minimized for $c_{j}=d_{j}=0$, which corresponds to
vanishing condensates $\sigma=\eta=0$.

\begin{figure}[tbh]
\centering
\includegraphics[width=0.49\textwidth]{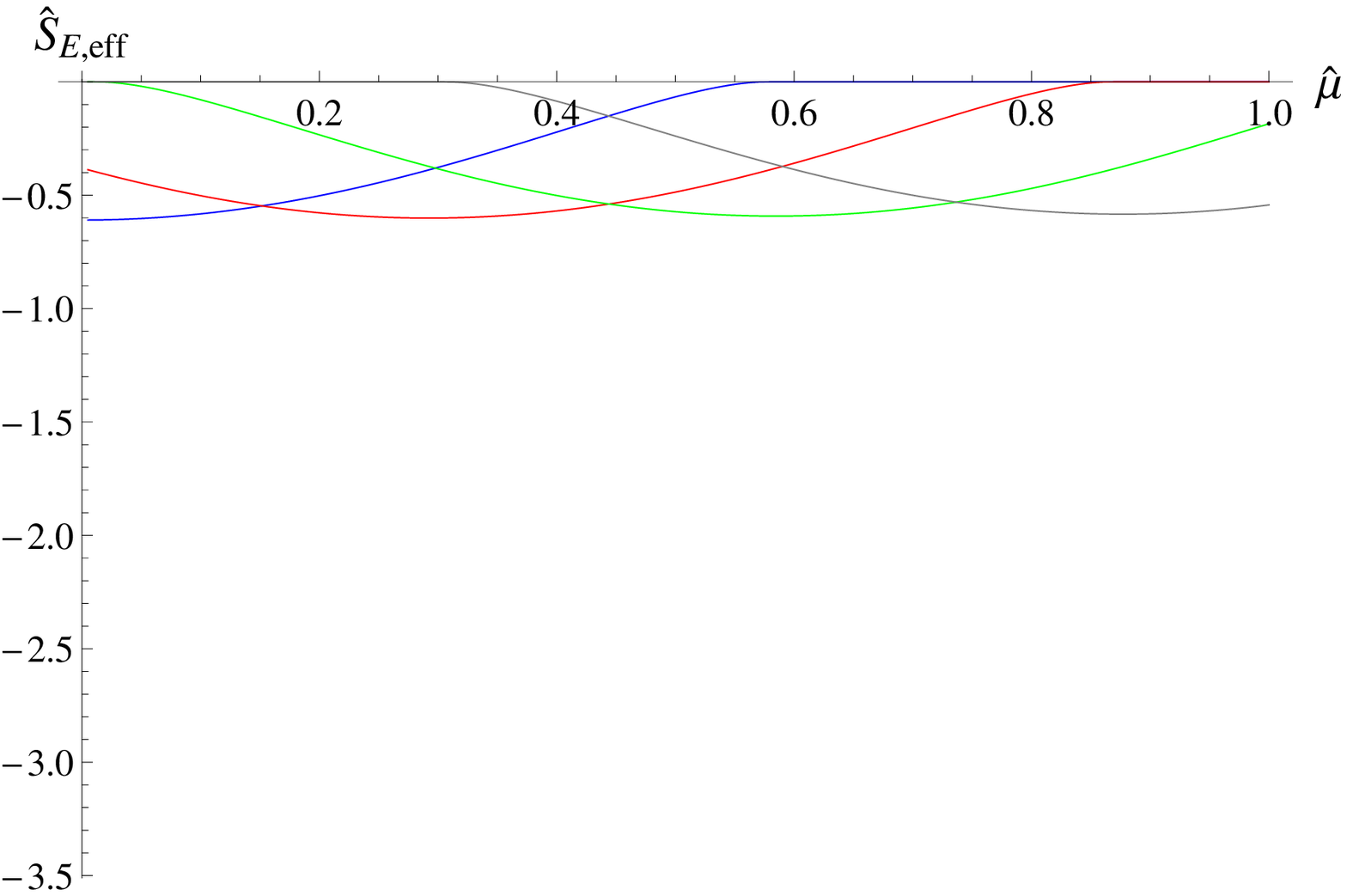}
\includegraphics[width=0.49\textwidth]{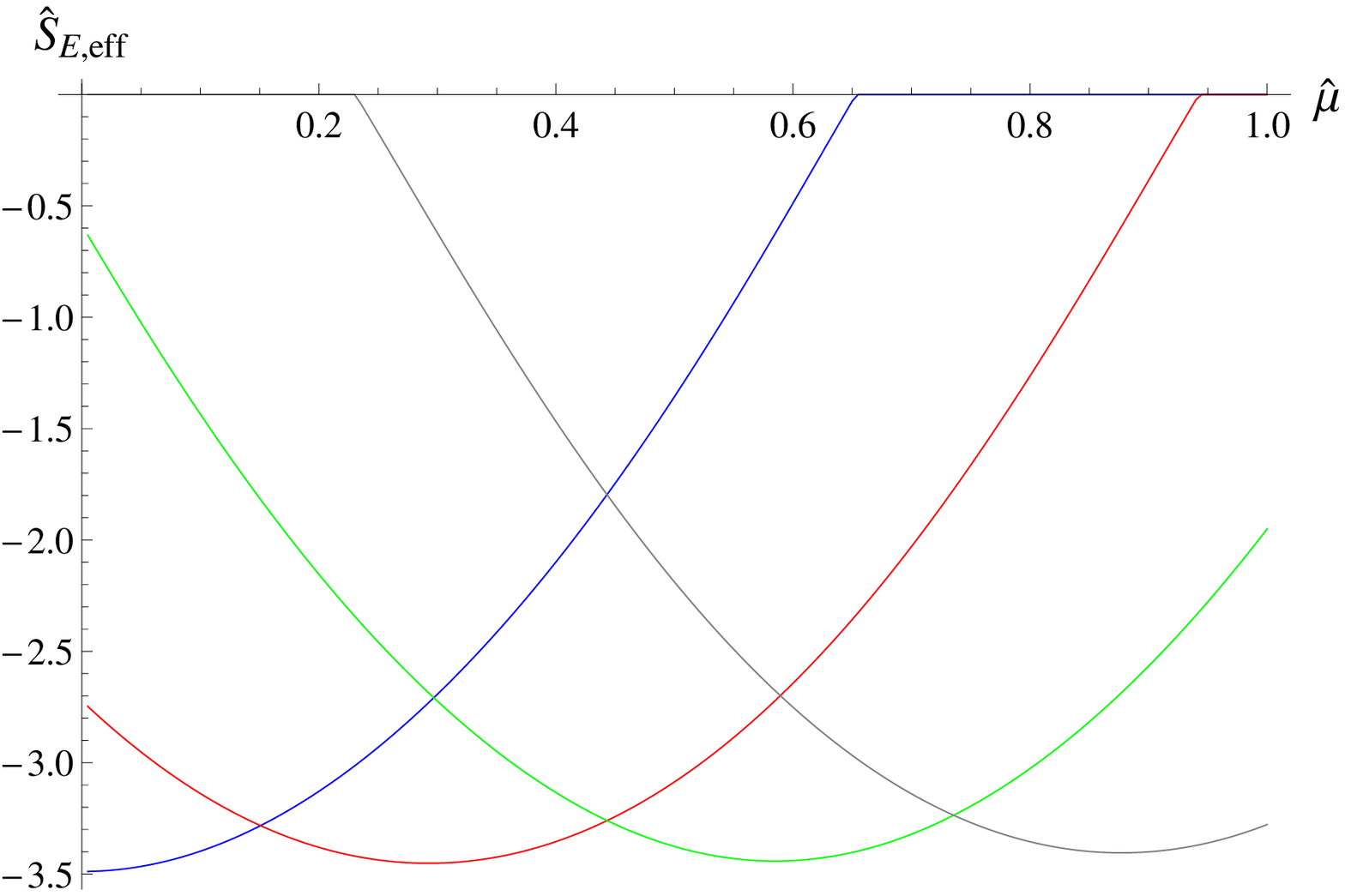}\caption{$\chi$GN
model. Local minimal values of $S_{E,\text{eff}}$ as functions of the chemical
potential $\mu$ for $m = 0$ (blue), $m = 1$ (red), $m = 2$ (green) and $m = 3$
(gray) and two different temperatures $\hat{T} = 0.378$ ($N_{0} = 24$, left
panel) and $\hat{T} = 0.189$ ($N_{0} = 48$, right panel).}%
\label{Fig:GN_CDW}%
\end{figure}

\begin{figure}[tbh]
\centering
\includegraphics[width=0.49\textwidth]{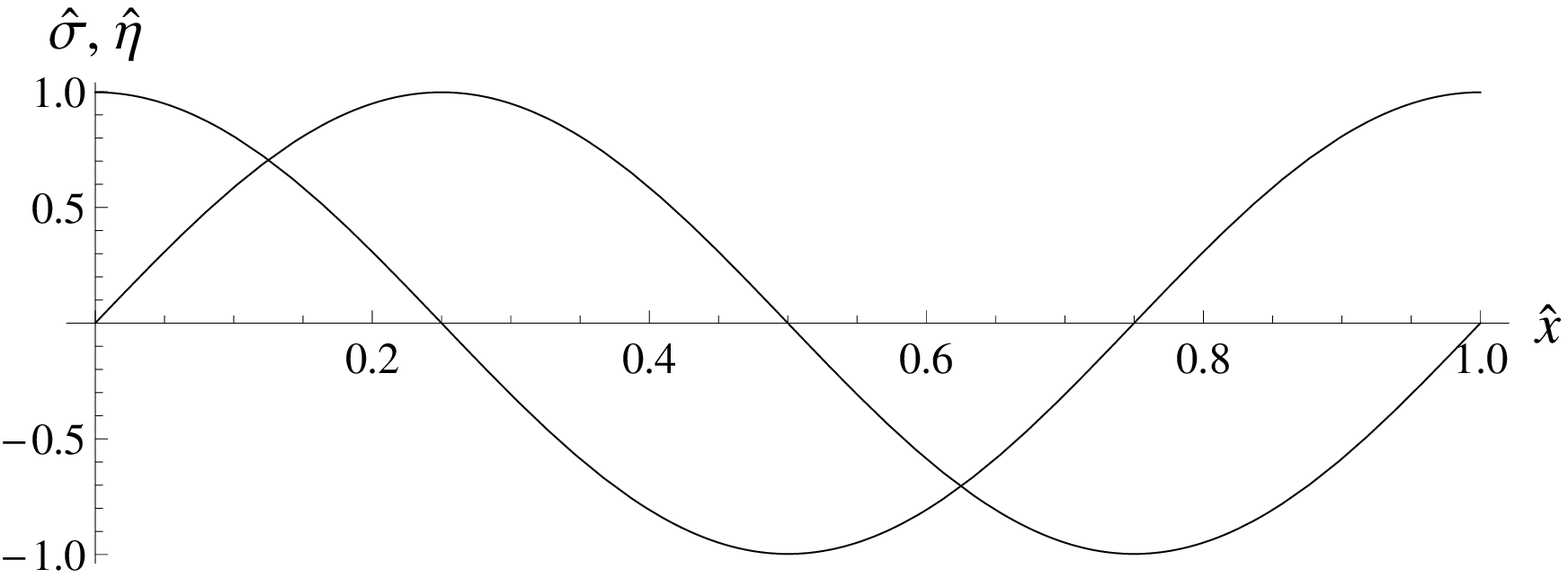}
\includegraphics[width=0.49\textwidth]{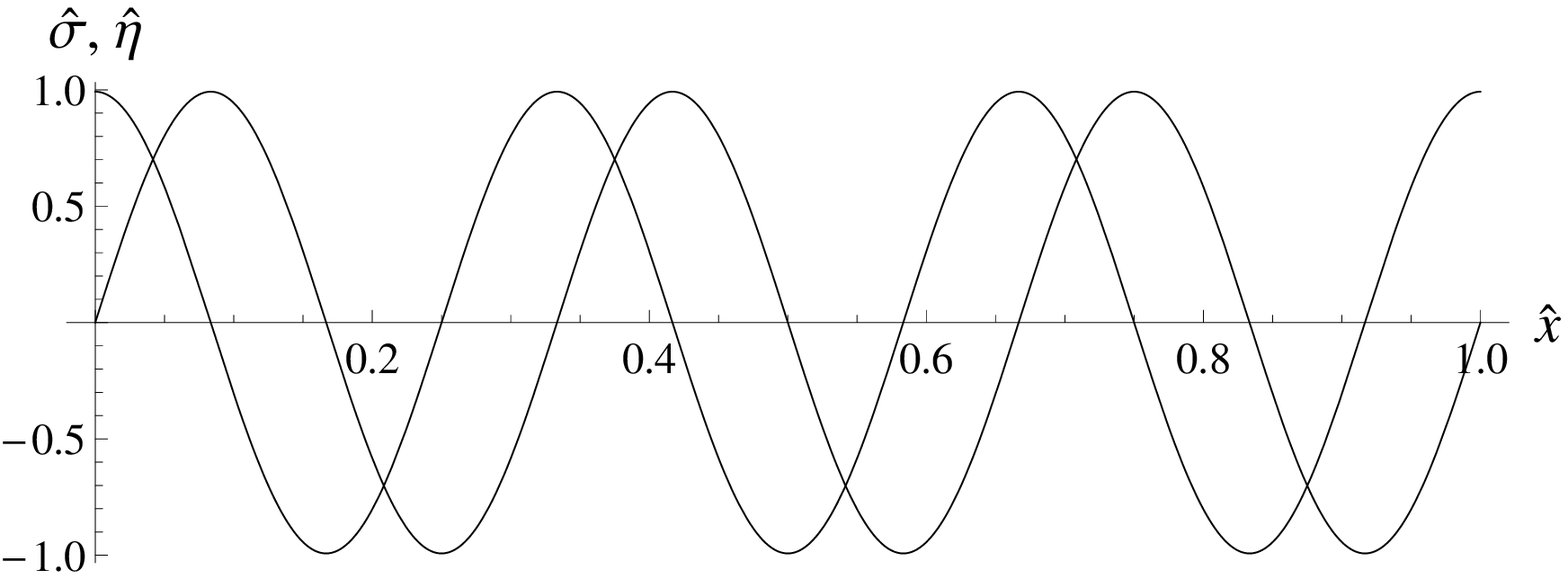}\caption{$\chi$GN model.
Resulting chiral CDWs for $\hat{T} = 0.095 < \hat{T}_{c}$ ($N_{0} = 96$) and
$\hat{\mu} = 0.295$ (left panel) and $\hat{\mu} = 0.875$ (right panel).}%
\label{Fig:GN_CDW_e}%
\end{figure}

For $T < T_{c}$ we find several local minima of $S_{E,\text{eff}}$, which are
given by $c_{m} = \pm i d_{m} \neq0$ for a single mode $m$, while $c_{j} =
d_{j} = 0$ for all other modes, i.e., $j \neq m$. The condensates $\sigma$ and
$\eta$ are harmonic functions, i.e., CDWs, with the same amplitude, but with a
relative phase shift $\pm\pi/2$, implying $\Sigma= \text{constant}$
[cf.\ Eq.\ (\ref{EQN799})]. The minimal values of $S_{E,\text{eff}}$ are
plotted in FIG.~\ref{Fig:GN_CDW} as functions of the chemical potential $\mu$
for $m = 0,1,2,3$ and two different temperatures $\hat{T} = 0.378, 0.189$
($N_{0} = 24, 48$, while $N_{00} = N_{1} = 96$ and $M = 10$). For $\mu= 0$ the
absolute minimum of $S_{E,\text{eff}}$ corresponds to $m=0$, i.e., $\sigma,
\eta= \text{const}$. The wavelength is proportional to $\mu$, i.e., for
increasing $\mu$ the absolute minimum of $S_{E,\text{eff}}$ corresponds to
larger and larger $m > 0$. Two examples of the resulting CDWs ($\hat{T} =
0.095 < \hat{T}_{c}$ [$N_{0} = 96$] and $\hat{\mu} = 0.295, 0.875$) are shown
in FIG.~\ref{Fig:GN_CDW_e}.

The resulting phase diagram is, therefore, quite different from the phase
diagram of the GN model: there are only two phases, for $T < T_{c}$ a CDW, and
for $T > T_{c}$ chiral symmetry is restored. This is in agreement with
analytically known results \cite{Schon:2000he, Basar:2009fg}.


\section{\label{SEC_NJL2}Phase diagram of the NJL$_{2}$ model}

Again the technical steps needed to compute the phase diagram closely parallel
those discussed for the GN model and the $\chi$GN model. This time four real
scalar fields $\sigma$ and $\pi_{j}$, $j=1,2,3$ are required, where
\begin{equation}
Z=\int D\sigma\bigg(\prod_{j=1}^{3}D\pi_{j}\bigg)e^{-S_{E,\text{eff}}%
[\sigma,\pi_{j}]}\quad,\quad S_{E,\text{eff}}[\sigma,\pi_{j}]=N\int
d^{2}x\,\bigg\{\frac{1}{2\lambda}\bigg(\sigma^{2}+\sum_{j=1}^{3}\pi_{j}%
^{2}\bigg)-\frac{1}{2}\ln[\det(Q^{\dagger}Q)]\bigg\}\;,
\end{equation}
with
\begin{equation}
Q=\gamma_{\mu}\partial_{\mu}+\gamma_{0}\mu+\sigma+\imath\gamma_{5}\sum
_{j=1}^{3}\tau_{j}\pi_{j}\;.
\end{equation}
$\sigma$ and $\pi_{j}$ as well as $\det(Q^{\dagger}Q)$ are then finite-mode
regularized as done in Eqs.\ (\ref{EQN751}), (\ref{EQN607}), and
(\ref{EQN608}) and Sec.~\ref{SEC470}, respectively. Due to the invariance of
$S_{E,\text{eff}}$ with respect to $(\sigma,\pi_{1},\pi_{2},\pi_{3}%
)\rightarrow R(\sigma,\pi_{1},\pi_{2},\pi_{3})$ with $R\in\text{O(4)}$,
dimensionful quantities are expressed in units of
\begin{equation}
\Sigma_{0}=\Sigma\Big|_{T=0,\mu=0}\quad,\quad\Sigma=\bigg(\sigma^{2}%
+\sum_{j=1}^{3}\pi_{j}^{2}\bigg)^{1/2}\;,
\end{equation}
and denoted by a hat $\hat{\phantom{x}}$.

We have studied the phase diagram of the NJL$_{2}$ model using $M = 4$ modes
for the condensates and $N_{00} = N_{1} = 72$ modes for the fermionic
determinant. For any temperature and chemical potential the four condensates
are proportional to each other, i.e., $\sigma\propto\pi_{1} \propto\pi_{2}
\propto\pi_{3} \propto\Sigma$ and also proportional to the chiral condensate
$\sigma$ of the GN model. Consequently, we obtain exactly the same phase
diagram for the NJL$_{2}$ model as for the GN model, which is shown in
FIG.~\ref{Fig:GN_pd_192}. These findings extend existing results, where only a
CDW [$\sigma(x) \propto\cos(2 b x)$ and $\pi_{3}(x) \propto\sin(2 b x) $] has
been considered \cite{Ebert:2011rg}. Our results show that an inhomogeneous
phase is indeed present at larger $\mu$ and not too large $T$. However, this
inhomogeneous phase exhibits solitonic structures and not a CDW. This result
is similar to those found for the NJL model in $3+1$ dimensions, as discussed
e.g.\ in Sec.~\ref{SEC_NJL} or Ref.\ \cite{Carignano:2011gr}.


\section{\label{SEC_NJL}Phase diagram of the NJL model}

We investigate the phase diagram of the NJL model at nonzero temperature and
density under the assumption that only the chiral condensate $\sigma$ defined
in Eq.\ (\ref{condnjl}) condenses. Thus, we do not take into account
condensation of the pion-like field combinations $\bar{\psi}_{j,f}\vec{\tau
}\imath\gamma_{5}\psi_{j,f}$ appearing e.g.\ in Eq.\ (\ref{njllag}). The
chiral condensate $\sigma$ is, in general, a function of the three spatial
coordinates $\vec{x}=(x_{1},x_{2},x_{3})$, i.e., $\sigma=\sigma(\vec{x})$.
However, previous investigations based on special ans\"{a}tze have shown that
modulations in more than one dimension are not favoured energetically
\cite{Carignano:2014jla,Buballa:2014tba}. Thus, for the sake of simplicity in
the following we assume that $\sigma$ depends only on one of the three spatial
coordinates, e.g.\ $x_{3}$, i.e.,
\begin{equation}
\hat{\sigma}=\hat{\sigma}(\hat{x}_{3})=\frac{\sigma(x_{3})}{\sigma_{0}}%
=\sum_{m=-M}^{M}c_{m}\frac{e^{-\imath\hat{p}\hat{x}_{3}}}{\sqrt{\hat{L}_{3}}%
}\quad,\quad c_{+m}=(c_{-m})^{\ast}\quad,\quad\hat{p}=\frac{2\pi}{\hat{L}_{3}%
}m\;.
\end{equation}
Of course, a study of the NJL model in the context of the finite-mode approach
without this assumption is an interesting topic which we plan to investigate
in the future.

Proceeding as in Sec.~\ref{SEC559} for the GN model we obtain the partition
function of the NJL model in $3+1$ dimensions,
\begin{equation}
Z=\int D\sigma\,e^{-S_{E,\text{eff}}[\sigma]}\quad,\quad S_{E,\text{eff}%
}[\sigma]=N\int d^{4}x\,\bigg\{\frac{6G}{N^{2}}\sigma^{2}-\ln[\det(Q^{\dagger
}Q)]\bigg\}\;,
\end{equation}
where $Q=\gamma_{\mu}\partial_{\mu}+\gamma_{0}\mu+m_{0}^{\ast}$. The
Pauli-Villars regularization can be implemented by adding heavy fermions as
explained in Sec.~\ref{SEC776},
\begin{equation}
S_{E,\text{eff,PV}}[\sigma]=S_{E,\text{eff}}[\sigma]-N\int d^{4}x\,\sum
_{k=1}^{N_{\mathrm{PV}}}C_{k}\ln[\det(\tilde{Q}_{k}^{\dagger}\tilde{Q}%
_{k})]\;,
\end{equation}
with $\tilde{Q}_{k}=\gamma_{\mu}\partial_{\mu}+\gamma_{0}\mu+M_{k}.$

For a convenient comparison with the existing literature on the NJL model (and
in contrast to the previous sections, where we discussed 1+1 dimensional
models) we express our results in the following in units of GeV. To this end,
the two parameters of the NJL model, the coupling constant $G$ and the
Pauli-Villars energy scale $\Lambda_{\text{PV}}$, are fixed by requiring that
a certain effective mass $m_{0}^{\ast}$ is realized (we perform computations
for three different choices, $m_{0}^{\ast}\in\{ 250 \, \text{MeV} \, , \, 300
\, \text{MeV} \, , \, 350 \, \text{MeV} \}$) and that the pion decay constant
reproduces the correct value in the chiral limit, $f_{\pi}=88 \, \text{MeV}$
\cite{Carignano:2011gr,Nickel:2009wj}. [For the evaluation of the pion decay
constant in the framework of the NJL model, we refer to
Ref.\ \cite{Klevansky:1992qe}.]

\begin{figure}[tbh]
\centering
\includegraphics[width=0.49\textwidth]{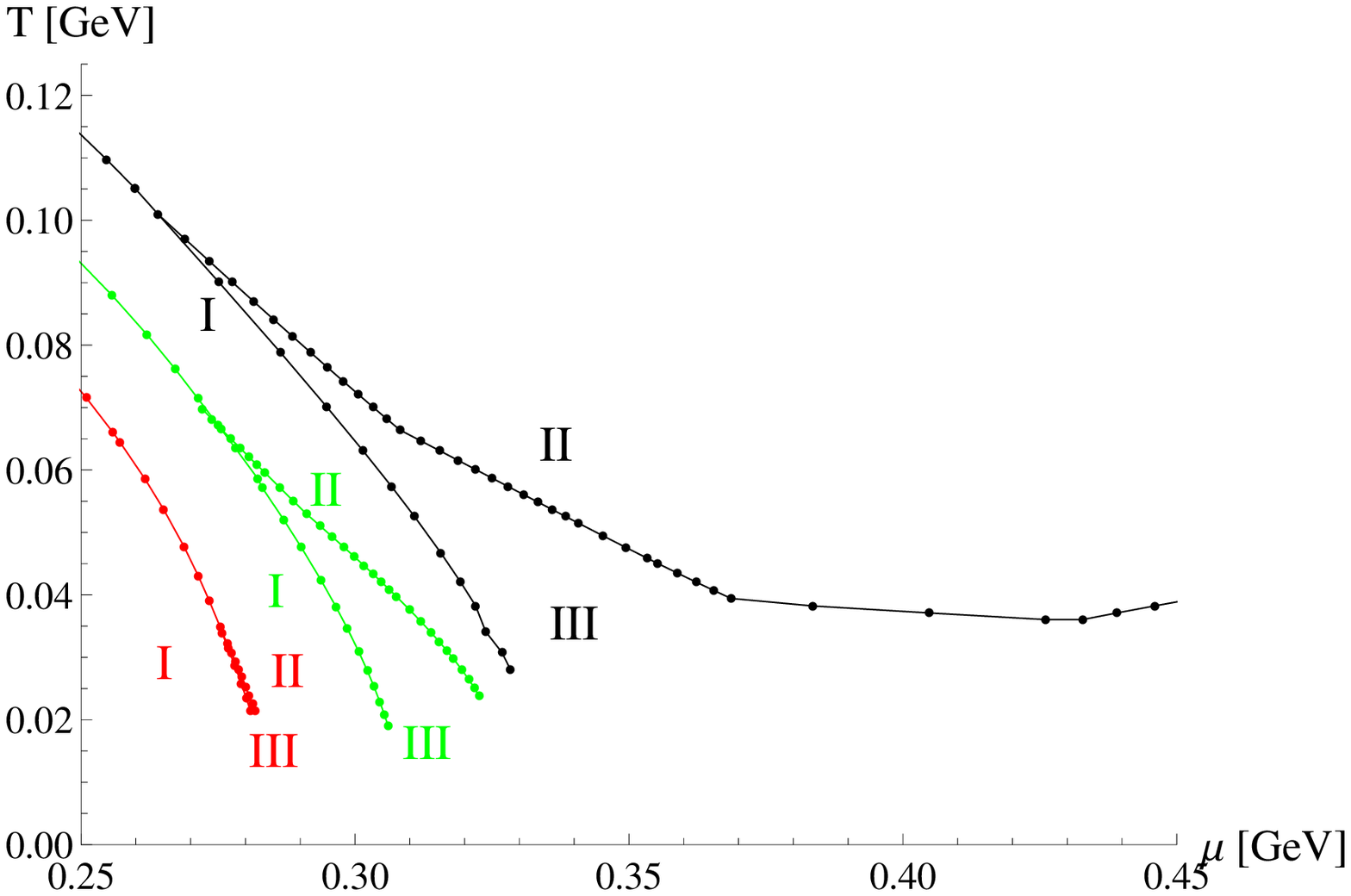}
\includegraphics[width=0.49\textwidth]{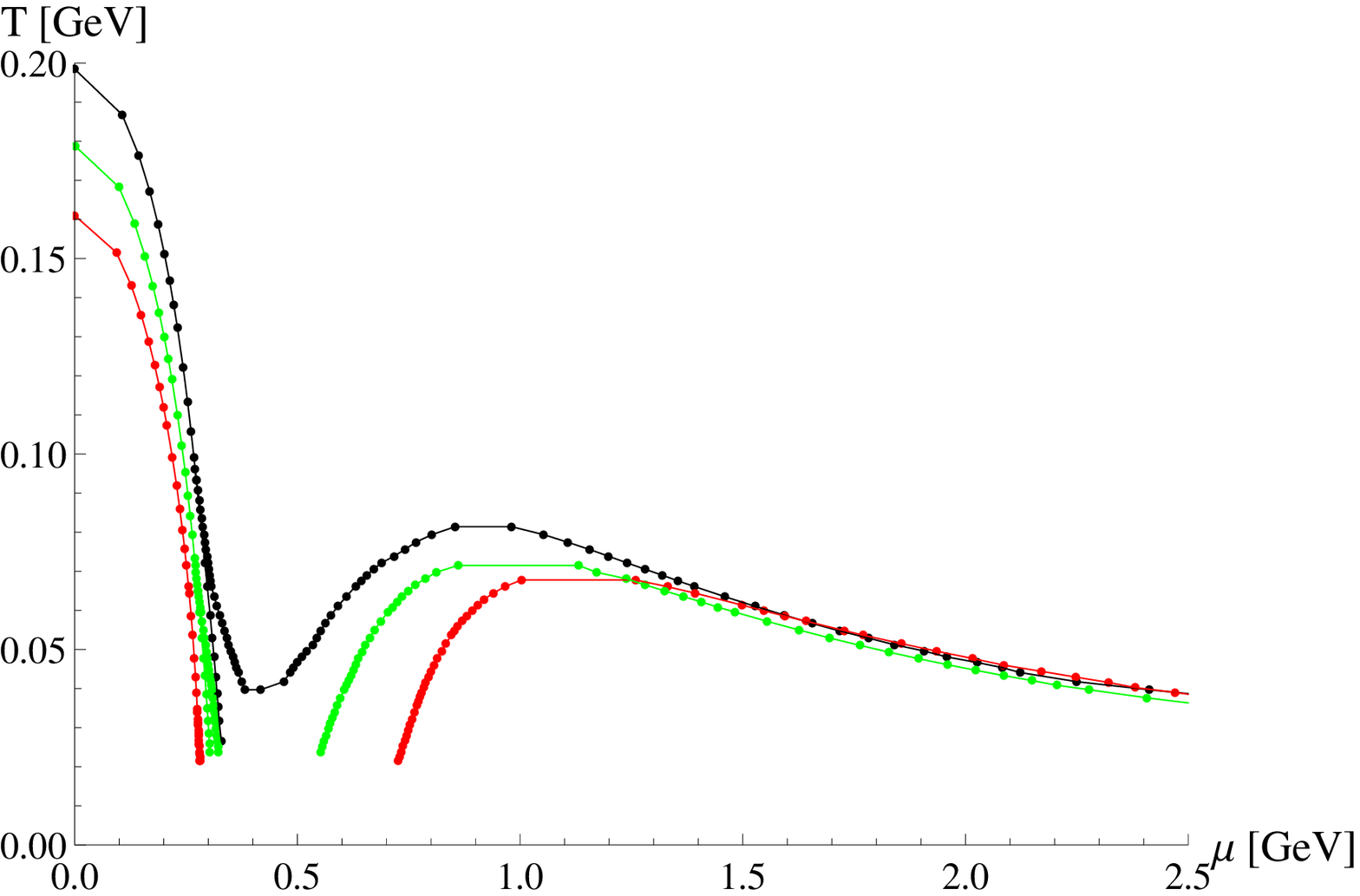}\caption{Phase
diagram of the NJL model for a spatially inhomogeneous condensate
$\sigma=\sigma(x_{3})$ for $N_{\text{PV}} = 2$, $M =5$, $N_{00} = N_{1} = 120$
and three different effective quark masses $m_{0}^{\ast}= 250$ (red),
$m_{0}^{\ast}= 300$ (green) and $m_{0}^{\ast}= 350 \, \text{MeV}$ (black).
There are three phases: (I) a homogeneous chirally broken phase $\sigma=
\text{constant} \neq0$, (II) a chirally restored phase $\sigma= 0$, (III) an
inhomogeneous phase (the left panel is a zoomed version of the right panel). }%
\label{Fig:NJL_reg3_comp}%
\end{figure}

The resulting phase diagrams for $N_{\text{PV}}=2$ additional heavy fermions,
effective quark masses $m_{0}^{\ast}\in\{250\,\text{MeV}\,,\,300\,\text{MeV}%
\,,\,350\,\text{MeV}\}$ and $M=5$ and $N_{00}=N_{1}=120$ modes are shown in
FIG.\ \ref{Fig:NJL_reg3_comp}. The corresponding Pauli-Villars cutoffs are
$\Lambda_{PV}\equiv\{736.8\,$MeV$,\,647.4\,$MeV$,\,608.7\,$MeV$\}$, the
maximum momentum used in the expansion of the chiral condensate is
approximately $(M/N)\hat{k}_{1}^{\text{cut}}\approx650\,\text{MeV}$, where
$\hat{k}_{1}^{\text{cut}}$ denotes the maximum momentum in the expansion of
the fermionic fields, see Eq.\ (\ref{cutoffk}). 

Let us discuss the role of the two energy scales $\Lambda_{PV}$ and $\hat
{k}_{1}^{\text{cut}}$ in somewhat greater detail. As mentioned already in
Sec.\ II.B, the NJL model is non-renormalizable. As a consequence, the NJL
model is only defined once the regularization has been fixed. Moreover, the
corresponding energy scale (the Pauli-Villars cutoff $\Lambda_{PV}$ in our
implementation of the NJL model) should be regarded as a physical parameter of
the NJL model, with crucial impact on all physical quantities. On the other
hand, our numerical approach contains also a cutoff due to the lattice
regularization in momentum space, i.e., due to using a finite number of modes,
see Eq.\ (\ref{cutoffk}). The cutoff $\hat{k}_{1}^{\text{cut}}$ is purely
technical and should be taken as large as possible. In particular, $\hat
{k}_{1}^{\text{cut}}$ should be much larger than $\Lambda_{PV}.$ In the
numerical calculations leading to Fig.\ 8 we have that $\hat{k}_{1}%
^{\text{cut}}/\Lambda_{PV}=25.4,$ thus $\hat{k}_{1}^{\text{cut}}$ is about
$16$ GeV. This is indeed a very large value that assures that for all
practical purposes the results do not depend on $\hat{k}_{1}^{\text{cut}}$.
Such a high value also assures that the continent of Fig.\ 8 is not an
artifact of our numerical calculation. We have also verified that relevant
quantities, such as the critical temperature, do not depend on $\hat{k}%
_{1}^{\text{cut}}$ once it is chosen sufficiently large.

Quite remarkably, these phase diagrams are similar to that obtained for the
1+1 dimensional GN model (cf.\ FIG. \ref{Fig:GN_pd_192}). This result suggests
that the 3+1 dimensional NJL model, which represents a non-renormalizable, but
in many aspects realistic chiral model of QCD, generates a phase diagram whose
most salient features can be understood in a simpler 1+1 dimensional field
theory. However, it should also be stressed that this result is obtained in
the specific case of the Pauli-Villars regularization and that the existence
of an inhomogeneous condensate depends on the chosen regularization scheme.

Independent of the concrete choice of the effective quark mass $m_{0}^{\ast}$
there is an inhomogeneous phase for large chemical potential $\mu$ and small
temperature $T$, termed \textquotedblleft continent\textquotedblright\ in
Ref.\ \cite{Carignano:2011gr}. However, at smaller $\mu$ the detailed shape of
the phase diagram depends on the value of the effective quark mass. While at
small $m_{0}^{\ast}$ an inhomogeneous \textquotedblleft
island\textquotedblright\ may be separated from the \textquotedblleft
continent\textquotedblright, at larger $m_{0}^{\ast}$ the island and the
continent merge. Note that our results in the region $\mu<0.4\,\text{GeV}$ are
in agreement with the findings of Ref.\ \cite{Nickel:2009wj}. At somewhat
larger $\mu$ they agree with the results of Ref.\ \cite{Carignano:2011gr},
although the outlines of the continent were not traced to very large $\mu$ in
that work. We find that, at even larger values of $\mu$, the transition
temperature between the chirally restored and the inhomogeneous phase
decreases with $\mu$, which is similar to the GN model.

We have repeated the study of the phase diagram of the NJL model for different
volumes and verified that the results are very stable upon changing the
volume. In particular, the form of the continent remains practically unchanged
when $N_{00}$ is modified. We have checked the cases $N_{00}=N_{1}=48,$ $72,$
$96,120$ and already for $N_{00}=N_{1}=48$ the curves look just as in Figs.\ 8
and 9. We have then selected for the plots the highest number of modes
($N_{00}=N_{1}=120$) with which we could perform our numerical study in a
reasonable computational time.

\begin{figure}[tbh]
\centering
\includegraphics[width=0.49\textwidth]{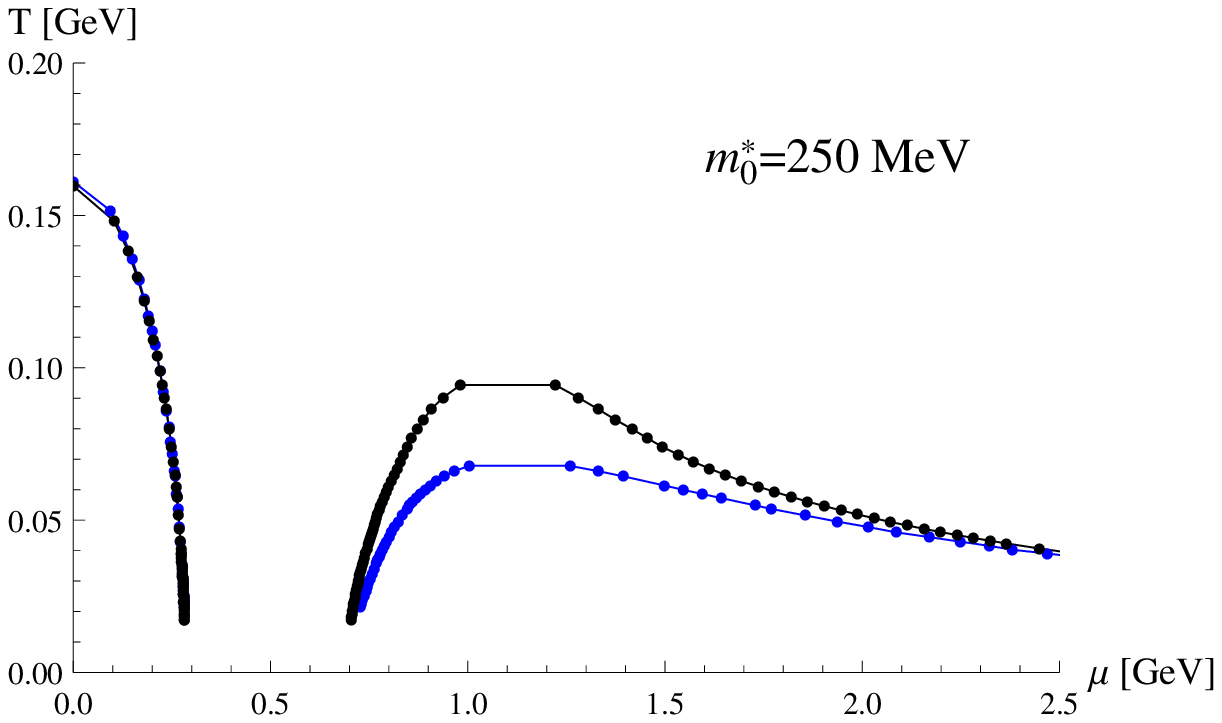}
\includegraphics[width=0.49\textwidth]{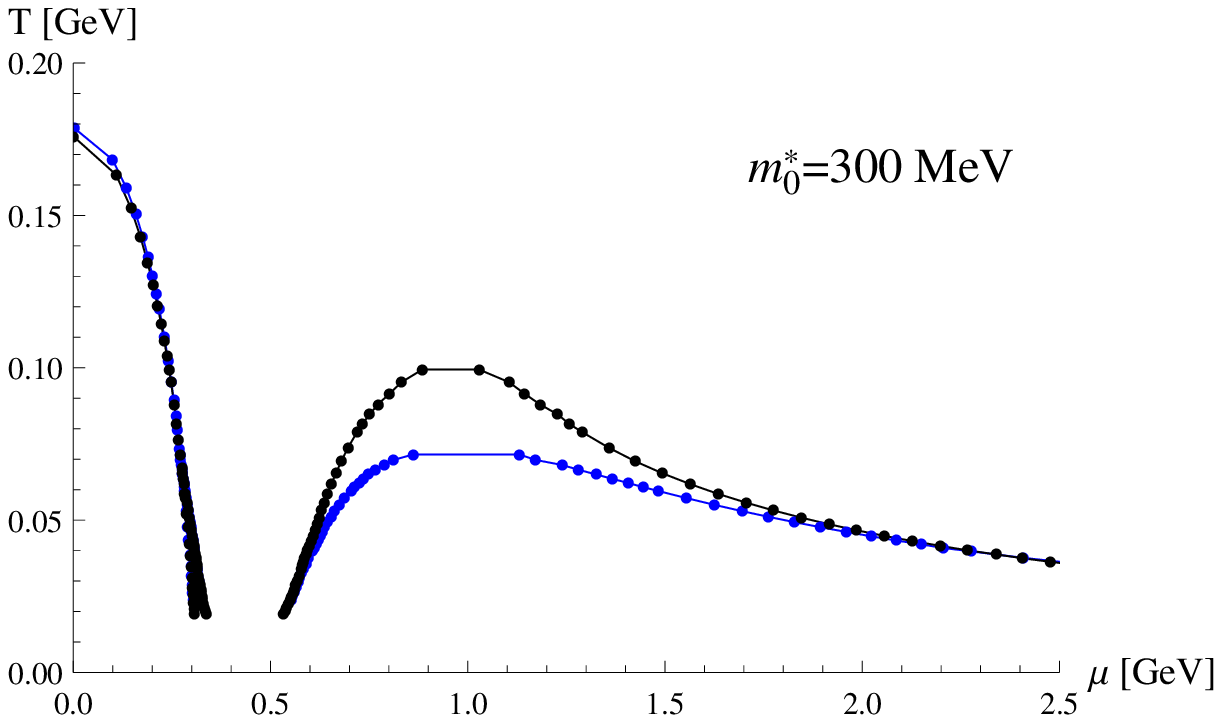}
\includegraphics[width=0.49\textwidth]{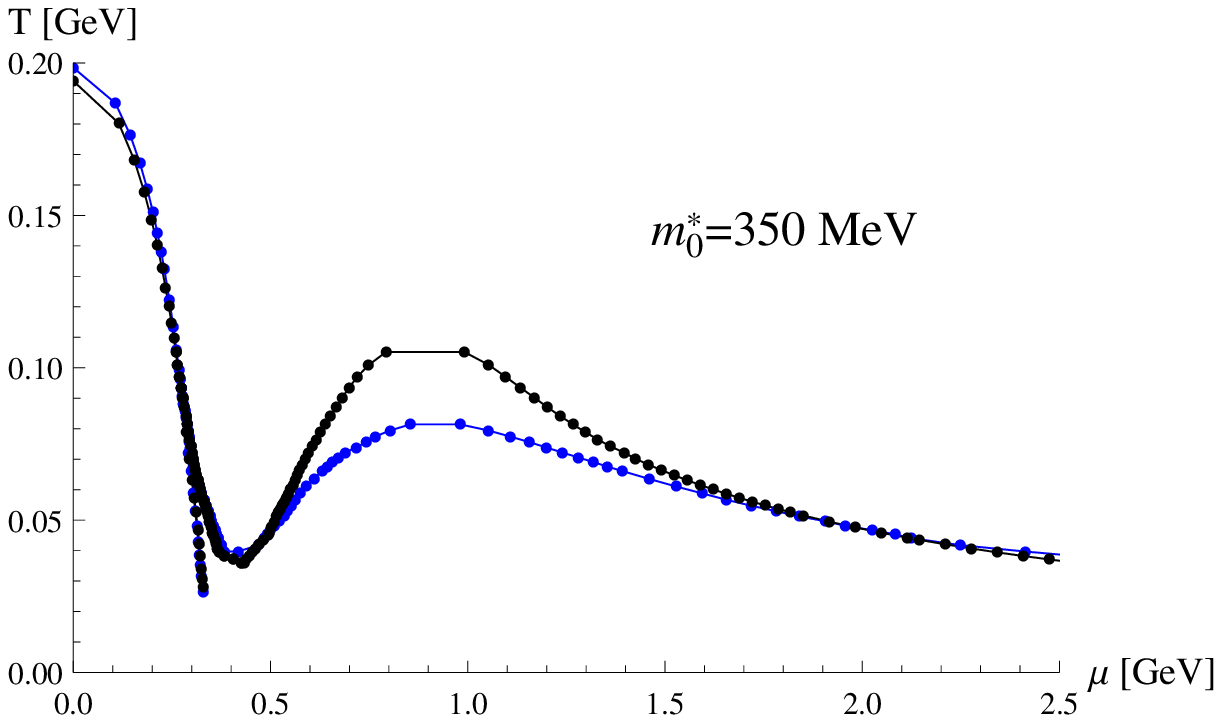}\caption{Phase
diagram of the NJL model for a spatially inhomogeneous condensate for
$N_{\text{PV}} = 2$ (blue) and $N_{\text{PV}} = 3$ (black) (further parameters
chosen as in FIG.\ \ref{Fig:NJL_reg3_comp}). }%
\label{Fig:NJL_reg3a5_comp}%
\end{figure}

We have also compared regularizations using $N_{\text{PV}}=2$ and
$N_{\text{PV}}=3$ additional heavy fermions
(cf.\ FIG.\ \ref{Fig:NJL_reg3a5_comp}). The effect of $N_{\text{PV}}$ on the
shape of the phase diagram is rather mild: at intermediate chemical potential
$\mu$ and temperature $T$ the inhomogeneous continent becomes somewhat larger
when using a larger number of regulators.

Although a detailed study of the order of phase transitions at the phase
boundaries is in principle straightforward, it would require a substantial
numerical effort, especially in the case of a second-order phase transition,
as expected for instance at the II-III boundary. We have, however, verified
that $\left\langle \hat{\sigma}^{2}\right\rangle $ smoothly approaches zero at
the II-III boundary. On the other hand, for small and constant temperature,
$\left\langle \hat{\sigma}^{2}\right\rangle $ is a slowly decreasing function
of the chemical potential $\mu.$

\section{\label{SEC4}Summary and outlook}

In this work we have used the finite-mode approach to investigate numerically
the emergence of inhomogeneous chiral condensation in effective quark models
in $1+1$ and $3+1$ dimensions. The main aim has been the determination of
inhomogeneous condensation in QCD-inspired models without using a specific ansatz.

We have shown that our method accurately reproduces well-known analytical
results concerning the phase diagram and the inhomogeneous condensation of
$1+1$ dimensional models, in particular the Gross-Neveu and chiral Gross-Neveu
model. By applying the approach to the NJL model in $3+1$ dimensions we could
reproduce previous results based on specific ans\"{a}tze for small chemical
potential. In addition to that we were able to show that the inhomogeneous
continent of Ref.\ \cite{Carignano:2011gr} extends to very high densities, but
not to arbitrarily large temperatures. Due to the fact that our results for
the NJL phase diagram differ from previous ones at high density, it would be
an interesting task for future studies to confirm or to falsify the presence
of the continent using different approaches.

It is interesting to note that the GN model and the NJL$_{2}$ model have the
same phase diagram, which in turn is also very similar to the NJL model in 1+3
dimensions. On the contrary, the phase diagram of the $\chi$GN model is
completely different, showing that a different flavor structure has an impact
on the phase diagram.

Our approach following Refs.\ \cite{Wagner:2007he,Wagner:2007av} is based on
expanding fermionic fields and condensates in terms of plane waves. In this
respect it is quite similar to existing general methods to compute
inhomogeneous condensates \cite{Nickel:2008ng,Carignano:2012sx,Cao:2015rea}.
In other aspects it is, however, quite different. For example our method is
based on a minimization of the effective action, not the Hamiltonian. A
detailed and direct comparison regarding the computational efficiency of our
method and those discussed in
Refs.\ \cite{Nickel:2008ng,Carignano:2012sx,Cao:2015rea} is difficult and
would require computations of the same quantities within the same model using
both types of methods. Since the basic idea of expanding in terms of plane
waves is the same, we expect both types of methods to perform on a
quantitatively similar level. Therefore, further studies would be necessary to
clarify why the shape of the \textquotedblleft continent\textquotedblright%
\ (see Fig.\ \ref{Fig:NJL_reg3a5_comp}) looks different in the two methods.

Note that within our approach it is straightforward to replace the plane waves
by another set of basis functions, as discussed and numerically demonstrated
in Refs.\ \cite{Wagner:2007he,Wagner:2007av}. A first and straightforward step
would be to still use a plane-wave expansion for the condensate, but to focus
exclusively on higher modes in regions where the condensate is expected to
exhibit strong oscillations (typically in inhomogeneous regions at large
chemical potential $\mu$). Using a small number of modes with suitably chosen
wave number might allow to explore the phase diagram quite extensively at
rather moderate computational cost. The choice of the wave numbers could even
be automated, i.e., included in the minimization procedure for the effective
action. Another possibility within our numerical framework is to study
specific ans\"{a}tze to identify a small set of highly relevant degrees of
freedom. This could not only provide certain physical insights, but also
reduce the computing time significantly. The latter might be of particular
importance, when studying two- or three-dimensional variations of the condensates.

In the future one could also apply the finite-mode approach to study the phase
diagram of purely hadronic theories such as the extended Linear Sigma Model
\cite{Parganlija:2012fy,Gallas:2009qp}. This approach is capable to correctly
describe the vacuum phenomenology as well as the nuclear matter ground state
properties \cite{Gallas:2011qp}. As already shown in
Ref.\ \cite{Heinz:2013hza}, an inhomogeneous condensate in the form of a
chiral density wave is favored with respect to a constant condensate at high
density ($\gtrsim2\rho_{0}$, where $\rho_{0}$ is the nuclear saturation
density). It is an open question whether other structures minimize the
effective potential even further. More generally, one could also apply the
finite-mode approach to quark-based sigma models, e.g.\ Refs.
\cite{Kovacs:2015vva,Tawfik:2014gga}.

Other interesting projects are the study of higher-dimensional modulations
beyond the ansatz used in Ref.\ \cite{Carignano:2012sx} and, since there is
also no limitation in the number of inhomogeneous fields in the finite-mode
approach, the study of interweaving chiral spirals \cite{Kojo:2011cn}. Further
effects at high densities such as inhomogeneous diquark condensation in $1+1$
as well as $3+1$ dimensions can also be taken into account
\cite{Ebert:2014woa}. Moreover, the models that we have studied were
investigated in the chiral limit only. Future work could thus include a
non-zero bare quark mass into the effective approaches.


\section*{Acknowledgments}

The authors thank M.\ Buballa, W.\ Broniowski, M.\ Sadzikowski, and
M.\ Praszalowicz for useful discussions.

M.W.\ acknowledges support by the Emmy Noether Programme of the DFG (German
Research Foundation), grant WA 3000/1-1.

This work was supported in part by the Helmholtz International Center for FAIR
within the framework of the LOEWE program launched by the State of Hesse.

Calculations on the LOEWE-CSC high-performance computer of Johann Wolfgang
Goethe-University Frankfurt am Main were conducted for this research. We would
like to thank HPC-Hessen, funded by the State Ministry of Higher Education,
Research and the Arts, for programming advice.



\begin{thebibliography}{99}
\bibitem {fn}Let $f(x)$ be an eigenfunction and $\alpha$ be the corresponding
eigenvalue of $Q$, i.e., $Qf(x)=(\gamma_{\mu}\partial_{\mu}+\gamma_{0}%
\mu+\sigma)f(x)=\alpha f(x)$. Then $(\gamma_{\mu}\partial_{\mu}+\gamma_{0}%
\mu+\sigma)f^{\ast}(x)=\alpha^{\ast}f^{\ast}(x)$, i.e., the eigenvalues
$\lambda$ are either real or form complex conjugate pairs. Therefore, $\det Q$
is real.
\end{thebibliography}
\end{document}